\pdfoutput=1

\documentclass[
  aps,
  physrev,
  reprint,
  superscriptaddress,
]{revtex4-2}

\usepackage{graphicx}
\usepackage{amsmath,amssymb}
\usepackage{bm}
\usepackage{hyperref}

\begin{document}

\title{Minimally Destructive Fast Imaging of Single Atoms in an Optical Tweezer Array with Coherent Excitation}

\author{Rei Yokoyama}
\thanks{These authors contributed equally to this work.}
\author{Takumi Kashimoto}
\thanks{These authors contributed equally to this work.}
\author{Kosuke Shibata}
\thanks{These authors contributed equally to this work.}
\author{Yuki Kawamura}
\author{Toshi Kusano}
\author{Chih-Han Yeh}
\author{Reiji Asano}
\affiliation{Department of Physics, Graduate School of Science, Kyoto University, Kyoto 606-8502, Japan}

\author{Yuma Nakamura}
\affiliation{Yaqumo Inc., 2-3-2 Marunouchi, Chiyoda-ku, Tokyo 100-0005, Japan}

\author{Tetsushi Takano}
\affiliation{Department of Physics, Graduate School of Science, Kyoto University, Kyoto 606-8502, Japan}
\affiliation{The Hakubi Center for Advanced Research, Kyoto University, Kyoto 606-8502, Japan}

\author{Yosuke Takasu}
\author{Yoshiro Takahashi}
\affiliation{Department of Physics, Graduate School of Science, Kyoto University, Kyoto 606-8502, Japan}
\altaffiliation[]{Contact author: yokoyama.rei.33w@st.kyoto-u.ac.jp}

\date{\today}

\begin{abstract}
Ultracold neutral atoms in an optical lattice and an optical tweezer array offer highly-controllable quantum many-body systems, utilized for various quantum science and technology such as quantum computing, quantum metrology, and quantum simulation.
By combining high-fidelity imaging of individual atoms, one can further enhance the capability of such experimental platforms as quantum gas microscopes, tweezer clocks, and tweezer-array-based quantum computers.
In this work, we propose minimally destructive single-atom imaging by deterministic coherent excitation of atoms with alternately applied $\pi$-pulses from counter-propagating directions, mitigating the fundamental heating effect associated with the stochastic absorption process.
Using ytterbium-174 atoms trapped in an optical tweezer array, we experimentally demonstrate fast and low-loss single-atom imaging with a discrimination fidelity of 99.89(5)~\%  and a survival probability of 98.80(44)~\%  in 17.6~\textmu s. 
Importantly, our scheme exhibits the lower heating rate, about half of that of the former scheme utilizing the incoherent excitation.
This fast and minimally destructive imaging scheme is beneficial for relaxing the requirement on the trap depth, thereby enabling scalable atom imaging across a wide range of quantum science platforms.
\end{abstract}

\maketitle

\section{\label{sec:introduction}Introduction}
A highly controllable system of trapped neutral atoms such as quantum gas microscopes~\cite{Bakr2009,Sherson2010,Gross2021} and optical tweezer arrays~\cite{Kim2016,Barredo2016,Endres2016,Kaufman2021} provides versatile experimental platforms for quantum computing~\cite{Saffman2010, Saffman2016,Henriet2020,Wintersperger2023,saffman2025}, quantum metrology~\cite{Madjarov2019,Norcia2019,Finkelstein2024,kaubruegger2025}, and quantum simulation~\cite{Bloch2012,Gross2017,Schfer2020,Browaeys2020,Gross2021}.
The performance of such platforms relies on high-fidelity imaging of individual trapped atoms.
The standard method is fluorescence imaging where atoms absorb photons from (near-)resonant probe beams and subsequently emit photons, which are collected with a high numerical aperture (NA) imaging system and detected by a camera to determine the occupation of each trap ~\cite{Ott2016}.  

While this provides high single-atom-discrimination fidelities above 99.9~\%, long imaging times, typically milliseconds or longer, are required to cool atoms and mitigate the recoil-heating effect inevitably introduced as a result of light absorption and emission process and thus keep atoms localized within the trapping sites during the imaging~\cite{Covey2019}. 
Therefore, the method imposes fundamental limitation for the experimental cycle rate, especially problematic for quantum computing, where stabilizer measurements must be frequently performed for error correction.
In fact, computational runtime estimated in  the recent resource estimation studies for neutral-atom quantum processors~\cite{Zhou2025, sunami2025, webster2026, cain2026, xue2026} assumes stabilizer measurement cycles on the order of 1~ms.
Reducing the imaging time from the millisecond regime to the microsecond regime is therefore essential for achieving a practically useful runtime in a large-scale fault-tolerant quantum computation.
Although high-intensity probe beams without active cooling allow for fast imaging below 1 ms, the induced heating leads to atom loss from the trap post-imaging~\cite{miranda2015}.

Important progress has been recently reported to simultaneously realize the high-imaging fidelity, low losses, and the short imaging times on the order of microseconds $\it{without}$ the cooling during the imaging process.
The key difference from the previous methods is in the probe pulse irradiation process where the short probe pulses are alternately applied from counter-propagating directions, 
rather than simultaneous irradiation from both directions.
This sequence thus avoids the possibilities of most detrimental heating process of absorption of photon from one direction and subsequent stimulated emission of photon from counter-propagating direction, successfully resulting in the fast, high-fidelity imaging of single atoms in an optical lattice~\cite{Su2025} and an atom tweezer array with high survival probability~\cite{Falconi2025}. 
Note that this sequence still relies on the incoherent excitation process of probe beams. Therefore, in addition to the recoil heating associated with the spontaneous emission, it inherently suffers from an equal amount of heating contribution stemming from the stochastic nature of the absorption process.

Here in this work, we propose $\it{minimally}$ destructive single-atom imaging without active cooling, enabled by repetitive deterministic coherent excitation with a $\pi$-pulse and subsequent waiting time, thereby ideally halving the total recoil heating by mitigating the fundamental heating effect associated with the stochastic absorption process. 
We demonstrate this method with an optical tweezer array of ytterbium-174 ($^{174}$Yb) atoms, which offers an advantage of the existence of the excited $^1\text{P}_1$ state with a short radiative lifetime of 5.5 ns, useful for fast imaging. Furthermore, it exhibits reduced recoil heating due to the large atom mass with a recoil energy of $E_{\mathrm{r}} = (\hbar k)^2 / 2m $, where $\hbar$ is the reduced Planck constant, $k$ is the wavenumber of the light, and $m$ is the $^{174}$Yb atomic mass.
By detecting fluorescent photons over short timescales less than 20~\textmu s, we realize single-atom-discrimination fidelity of about 99.9~\% and survival probability of about 99~\% after a total of 1000 pulses. 
Importantly, with careful diagnostics, we successfully verified the lower heating nature of our scheme, about half in comparison with that of incoherent excitation scheme, which is consistent with the theoretical prediction.
The minimally destructive nature of our scheme is also promising for relaxing the requirement on the trap depth, thus leveraging atom imaging in a variety of quantum information processing.

\section{\label{sec:results}Results of Experiments}
Our experiment starts with a magneto-optical trapping (MOT) of $^{174}$Yb atoms using the $^1\text{S}_0 \leftrightarrow{} ^3\text{P}_1$ intercombination transition at a wavelength of 556 nm, 
followed by stochastic loading of atoms into a $6\times6$ two-dimensional array of optical tweezers formed by a laser light beam at 532 nm and a waist of 580 nm. The filling fraction is almost 50~\% through light-assisted collision process with the MOT beams.

\subsection{\label{sec:scheme}\textbf{Coherent Excitation Scheme of Imaging}}
\begin{figure*}[t]
  \centering
  \includegraphics[width=0.9\linewidth]{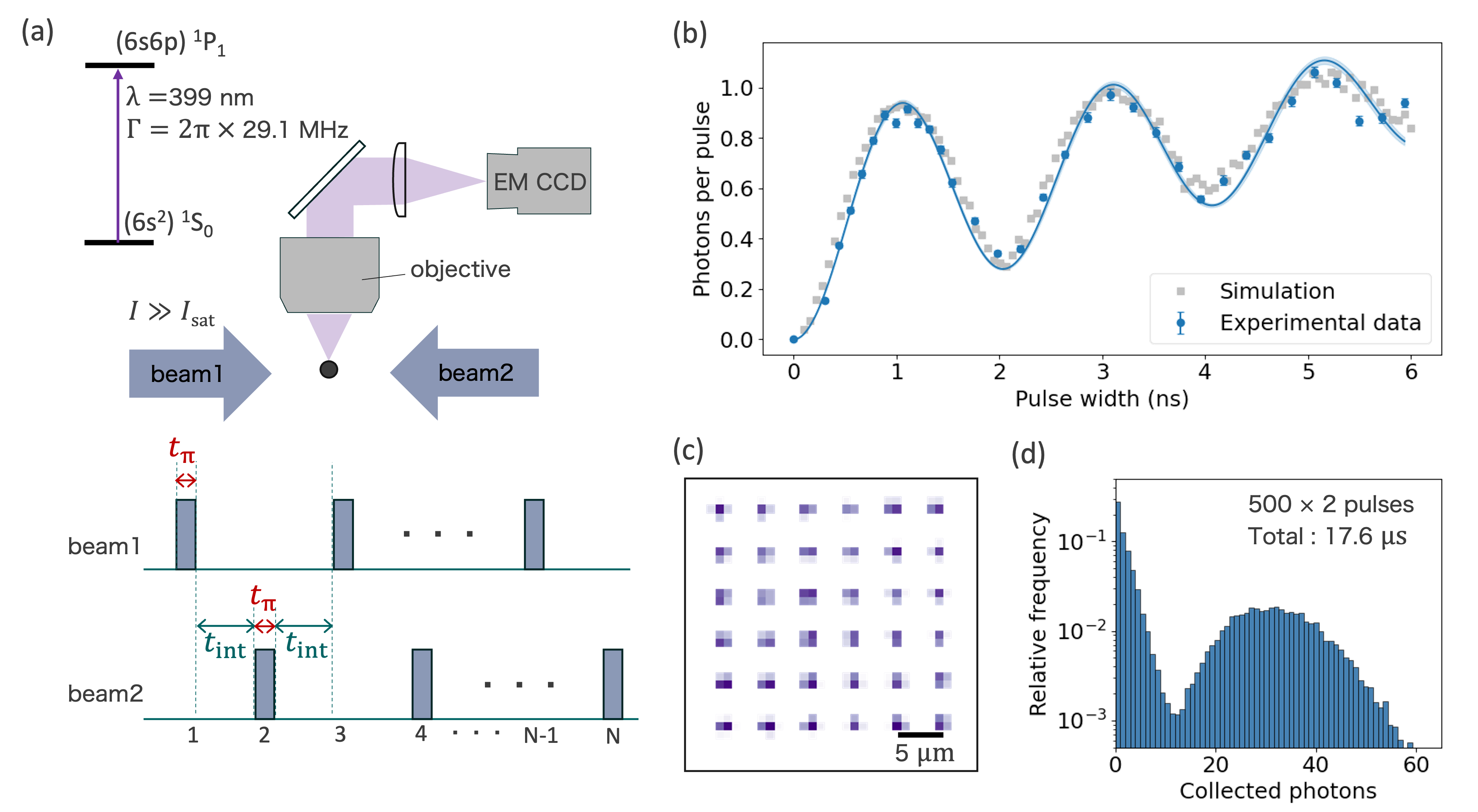}
  \caption{Imaging of a $^{174}$Yb atom array with coherent excitation. 
  (a) Schematic of the experimental setup and pulse sequence. An array of $^{174}$Yb atoms is addressed by a series of $\pi$-pulses resonant with the $^1\text{S}_0 \leftrightarrow{} ^1\text{P}_1$ 399 nm transition at an intensity $I \gg I_{\text{sat}}$ at intervals of $t_\mathrm{int}$. The spontaneously emitted photons are collected by a high NA = 0.6 and imaged onto an EM-CCD camera. Relevant energy levels and transition of $^{174}$Yb are shown on the upper left. 
  (b) Observation of Rabi oscillations via fluorescence detection as a function of pulse width. The experimental data are represented by circles and the error bars are the 1-$\sigma$ standard error of the mean. The fit of the data with the semiclassical analysis of Eq.~(\ref{eq:fluorescnce_rabi}) is shown by a solid line. The result of the Monte Carlo simulation is shown as grey squares.  
  (c) Average image over 300 realizations of a 6$\times$6 two-dimensional tweezer array of $^{174}$Yb atoms.
  (d) Histogram of the integrated photon counts within the region of interest for 1000 pulses, showing clear bimodal separation between the background and single-atom signals.}
  \label{fig:setup}
\end{figure*}

Figure~\ref{fig:setup}(a) shows the schematic illustration of our imaging with coherent excitation. Counter-propagating and alternated $\pi$-pulses resonant with the $^1\text{S}_0 \leftrightarrow{} ^1\text{P}_1$ transition at a wavelength of 399 nm are applied with a certain time interval to atoms trapped in the optical tweezer array.
The probe beams propagate along the horizontal plane, and their
polarization is aligned also in the horizontal plane and is parallel to a magnetic field to drive the $\pi$ excitation.
Photons spontaneously emitted from the atoms are collected by an objective lens with a NA of 0.6 and are focused and detected with an electron-multiplying-charge-coupled-device (EM-CCD) camera (iXon-Ultra897, Andor).

The distinctive feature of the coherent excitation scheme is the elimination of stochastic heating associated with photon absorption.
When we consider the absorption process, not only are the momentum kicks cancelled through alternated and counter-propagating $\pi$-pulses, resulting in the null average value, but the momentum variance is also eliminated, thereby nullifying the heating contribution from absorption.
This is due to the fact that, in the fully coherent limit, the number of absorbed photons is fixed to the number of pulses.
Heating is, thus, ideally induced only by the randomness of spontaneous emission direction. 
This is in contrast with the former implementation of incoherent scheme~\cite{Falconi2025} where both absorption and emission processes contribute equally to the stochastic energy gain.

A key for successful implementation of fast imaging with coherent $\pi$-pulse excitation is the generation of short and intense probe pulses. 
Specifically, the temporal width of the $\pi$-pulses should be shorter than the excited $^1\text{P}_1$ state lifetime of 5.5 ns to achieve coherent condition, requiring the associated Rabi frequency of several hundreds of MHz. This corresponds to the ultraviolet probe intensity of $10^4$ mW/cm$^2$ or $10^2$ $I_\text{sat}$, which is technically demanding.
We generate such probe light beam by pulsing 798 nm laser with a 40 GHz electro-optic (EO) intensity modulator (LNX7840A, Thorlabs Inc.) and doubling the laser frequency with a fiber-coupled, single-path MgO:PPLN waveguide second harmonic generation (SHG) module (RSH-M03999-P85L45AB0-KH, AdvR Inc.).
The pulse width of the 399 nm probe light beam is controlled by a fast pulse generator (PG-1072, Active Technologies), which can produce a pulse as short as 300 ps.
The details are described in Appendix~\ref{app:optical setp}.

Coherent excitation for the $^1\text{S}_0 \leftrightarrow{} ^1\text{P}_1$ probe transition, the most crucial building block of the proposed imaging scheme, is confirmed by the successful observation of a Rabi oscillation.
Figure~\ref{fig:setup}(b) shows the detected photon counts as a function of probe pulse width with a total of 800 excitation pulses at a fixed repetition period of 17.6 ns.
The detected photon counts shows a clear oscillatory behavior as the evidence of the coherent excitation.
From this result, we can determine the $\pi$-pulse time $t_\pi$ as 1.1 ns, which is well shorter than the $^1\text{P}_1$ lifetime of 5.5 ns.
One can see that the detected photon counts also show gradual increase as the pulse width increases. 
Note that the similar behavior was reported in a quantum dot system~\cite{Schaibley2013}.

We can quantitatively explain the observed behavior by a semiclassical analysis where the photon counts $S$ for the pulse width of $\tau$ can be written (except the overall scaling factor) as
\begin{equation}\label{eq:fluorescnce_rabi}
    S = \gamma_1 \int_{0}^{\tau}\rho_{ee}(t)dt + \rho_{ee}(\tau).
\end{equation}
where the first term corresponds to the spontaneous emission events during the pulse duration and the second the excited-state population at the end of the pulse which contributes to the spontaneous emission after the pulse irradiation.
Here the excited-state population $\rho_{ee}(t)$ is given by~\cite{Schaibley2013},
\begin{align}
    \rho_{ee}(t) =& \frac{\Omega^2/2}{\Omega^2+\gamma_1\gamma_2 } \nonumber\\ 
     &\times\left[ 1- \left( \cos(\Lambda t) + \frac{\gamma_1+\gamma_2}{2\Lambda}\sin(\Lambda t) \right) e^{-\frac{\gamma_1+\gamma_2}{2}t} \right],
\end{align}
where 
$\Omega$ is the Rabi frequency, $\gamma_1:=1/T_1$ is the excited state decay rate, $\gamma_2$ is the decoherence rate, and
\begin{equation}
    \Lambda = \sqrt{\Omega^2 -\frac{(\gamma_1 - \gamma_2)^2}{4}}.
\end{equation}
The derivation assumes the fluorescence process by each pulse is independent, meaning that the atom is in the ground state at the beginning of the pulse.
By straightforward calculation, the integral in Eq. (\ref{eq:fluorescnce_rabi}) is found to be 
\begin{equation}
    \frac{\Omega^2/2}{\Omega^2+\gamma_1\gamma_2 }
       \left( \tau - \frac{\sin(\Lambda \tau)}{\Lambda} e^{-\frac{\gamma_1+\gamma_2}{2}\tau} \right).
\end{equation} 
We obtain $\Omega = 2\pi \times 490.0(8)$ MHz from fitting the experimental data with 
Eq. (\ref{eq:fluorescnce_rabi}) with $1/\gamma_2$ fixed to $2/\gamma_1$.

We also perform a Monte Carlo simulation using QuTiP~\cite{qutip2}.
This approach incorporates the coherent evolution between the $^1\text{S}_0$ and $^1\text{P}_1$ states, driven by the periodic pulse sequence, while treats the spontaneous emission as a stochastic quantum jump process via the collapse operator $C_1 = \sqrt{\gamma_1} \sigma^-$, where $\sigma^- = |^1\text{S}_0\rangle\langle^1\text{P}_1|$ is the atomic lowering operator. 
We count jump events as the number of spontaneous emission events.
We set $\Omega=2\pi \times490$ MHz, pulse repetition rate fixed to 17.6 ns regardless of pulse length $\tau$, and total pulse number of 100 in the simulation. 
The simulation also reproduces the experimental result, as shown by grey squares in Fig.~\ref{fig:setup}(b).

\subsection{\label{sec : parameter dependence}\textbf{Fast Imaging with Coherent Excitation}}
Imaging is performed with counter-propagating and alternated 1.1 ns-long $\pi$-pulses, short enough for the coherent excitation scheme. 
A typical averaged image is shown in Fig.~\ref{fig:setup}(c). 
The total pulse number of 1000 and time interval of $t_\mathrm{int}=16.5~\mathrm{ns}\approx3T_1$ between successive $\pi$-pulses result in high-fidelity single-atom detection.
As can be seen in Fig.~\ref{fig:setup}(d), we are able to well distinguish the filled sites from the background.
This is enabled with short imaging time of 17.6~\textmu s, more than two orders of magnitude improvement over the standard imaging time of several ms or longer, and is only 2.8 times longer than the shortest imaging time in the former implementation~\cite{Falconi2025}. 

To quantitatively evaluate the imaging performance, we employ a two-shot model-free analysis, which determines the fidelity and survival probability directly from the experimental photon-count distributions of two consecutive imaging shots~\cite{Norcia2018, Holman2026}. 
Using this approach, we achieve a discrimination fidelity of 99.89(5)~\% and a survival probability of 98.80(44)~\% after a total of 1000 pulses in a 2.27 mK deep trap.
This loss could be partially due to off-resonant scattering from the excited state caused by the 532~nm tweezer light, which contributes 0.58~\% using the scattering rate evaluated in Ref.~\cite{Falconi2025}.

\begin{figure*}[t]
  \centering
  \includegraphics[width=0.9\linewidth]{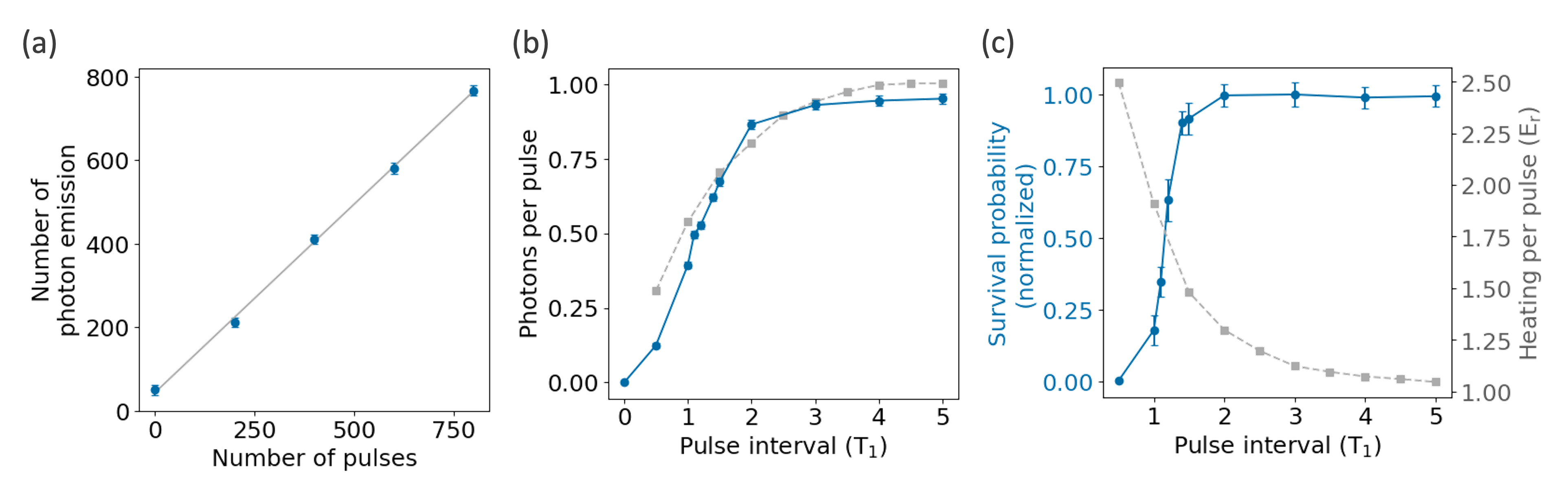}
  \caption{Systematic investigation of the coherent excitation scheme.
  (a)	Number of emitted photons measured as a function of the number of pulses. The detected photon counts are converted into the number of spontaneously emitted photons using the estimated photon collection efficiency of our imaging setup. The experimental data are represented by circles. A linear fit yields a emission rate of 0.90(2) photons per pulse.
  (b,c) Dependence of the emitted photon count per pulse (b) and normalized survival probability (c) on the pulse interval, using a sequence of 800 pulses. Suppression in both signals at shorter intervals indicates that stimulated emission occurs before spontaneous decay, leading to additional heating. The solid grey squares represent a Monte Carlo simulation, which shows a trend consistent with the experimental data.
  In (a-c), the error bars indicate the 1-$\sigma$ standard error of the mean.}
  \label{fig:parameter_dependence}
\end{figure*}

We then characterize the basic properties of our coherent excitation scheme. 
First, Fig.~\ref{fig:parameter_dependence}(a) shows that the number of emitted photons exhibits a linear dependence on the number of pulses applied at a pulse interval of 16.5 ns, yielding an average of 0.90(2) photons per pulse from the linear fit. 
Note that the total number of emitted photons is inferred from the detected photon counts using the photon collection efficiency of 3.4~\% of the imaging setup.
While the Eq.~(\ref{eq:fluorescnce_rabi}) predicts 1.03 photons per pulse, 
the finite repetition interval leaves a fraction of the atoms in the excited state, thereby reducing the number of photon emissions. We obtain 0.936(4) photons per pulse from the Monte Carlo simulation, which is reasonably close to the observed value of 0.90(2) considering the inherent uncertainty in the estimation of collection efficiency.

Next, we investigate the pulse interval dependence of the emitted photons per pulse and the survival probability to determine the optimal pulse interval.
This is important because, as pointed out in~\cite{Su2025}, the concurrent irradiation of two opposing probe beams can result in the coherent momentum gain due to coherent quantum walk in momentum subspace between spontaneous emission events.
In the proposed coherent excitation scheme, a similar additional heating phenomenon due to coherent momentum growth can occur, for example, a process in which a $\pi$-pulse from one direction excites the atom, followed by stimulated emission due to the second $\pi$-pulse from the opposite direction, prior to spontaneous emission. 
The consequence of this process is the acquisition of two-photon momentum in the same direction, whilst the spontaneous emission of the photon does not occur. 
To quantify the effect of this phenomenon, we examine the dependence of the survival probability and emitted photons per pulse on the pulse interval using a sequence of 800 counter-propagating and alternating $\pi$-pulses, as shown in Fig.~\ref{fig:parameter_dependence}(b) and (c), respectively. 
Considerable suppression both in survival probability and photon counts is observed at shorter pulse intervals, indicating that stimulated emission should occur prior to spontaneous decay and induce additional heating. 
For comparison, we perform Monte Carlo simulations for the photon counts and heating per pulse.
The simulation of photon counts shows a trend similar to the experimental result, as shown in Fig.~\ref{fig:parameter_dependence}(b) by grey squares.
The simulation of heating per pulse, represented by grey squares plotted with respect to the right axis in Fig.~\ref{fig:parameter_dependence}(c), reveals substantial heating at short pulse intervals, which is consistent with the lower survival probability experimentally observed.
Note that, when simulating heating per pulse, the definition of the state vector is extended to the tensor product of the atomic internal states and the momentum states to incorporate the coherent time evolution of the momentum states.

\subsection{\label{sec : suppressed heating}\textbf{Suppressed Heating with Coherent Excitation}}
The advantage of our scheme with coherent excitation over those utilizing the incoherent excitation is the smaller heating effect.  
Our careful diagnostics verifies that the heating rate in our scheme is about half of that with the incoherent excitation scheme, consistent with the theoretical prediction.

Note that, in this comparison of the heating effects, we take special care to suppress possible influence of very weak leakage light of the probe beams, typically on the order of 0.04 $I_{\text{sat}}$, mentioned in Appendix~\ref{app:optical setp}. This is done by detuning the laser frequency by $+$30 MHz, about one natural linewidth $\Gamma$/(2$\pi$) of probe transition, from the resonance.
This choice of the detuning reduces the scattering of atoms due to the leaked probe light beams by a factor of approximately 5, thus resulting in only about 12 scattering events, negligible in our temperature regime, while it does not so much affect the excitation by high-intensity probe pulses well beyond the saturation regime with Rabi frequencies much greater than $\Gamma$.
This is especially important for the coherent excitation scheme where most of the imaging time is spent for just waiting and thus the scattering of the very weak leackage light simply accumulates to a presumably non-negligible amount.

For fair comparison of the heating effects, we also carefully investigate the optimal conditions of the incoherent excitation scheme (see Appendix~\ref{app:incoherent}).
In particular, we choose the conditions which give nearly minimum heating effects (see Fig.~\ref{fig:incoherent}(c)).
Consequently, similar to the coherent excitation case, we employ trains of counter-propagating and alternated pulses~\cite{Bergschneider2018, Su2025, Falconi2025}, with less than 100 ns width with the intensity of 40 $I_{\text{sat}}$, corresponding to approximately 0.1 times of that in the coherent excitation scheme. The total photon counts becomes equivalent to that of 1000 pulses in the coherent excitation scheme by taking, for example, a total imaging duration of approximately 10~\textmu s consisting of hundred 100 ns pulses.
We also carefully consider the heating effect of weak leakage light, confirming the negligible contribution (see Appendix~\ref{app:incoherent} and Fig.~\ref{fig:incoherent}(b)).

Here, we specifically evaluate the $\it{heating~per~photon}$ of the coherent and incoherent excitation schemes.
By analyzing the linear dependence of temperatures measured by release-and-recapture method on the number of spontaneously emitted photons, the uncertainty associated with determining the initial temperature is eliminated.
Even so, evaluating absolute atomic temperatures involves several uncertainties, such as the effective recapture volume and other potential systematic effects~\cite{ref_coolingRR} in the release-and-recapture method. 
Furthermore, the uncertainty associated with the photon collection efficiency introduces a systematic uncertainty into the determined $\it{heating~per~photon}$.
To mitigate these unknown factors, we focus on the $\it{ratio}$  of the $\it{heating~per~photon}$ between the coherent and incoherent excitation schemes, thereby highlighting the difference between the two.

\begin{figure*}[t]
  \centering
  \includegraphics[width=0.9\linewidth]{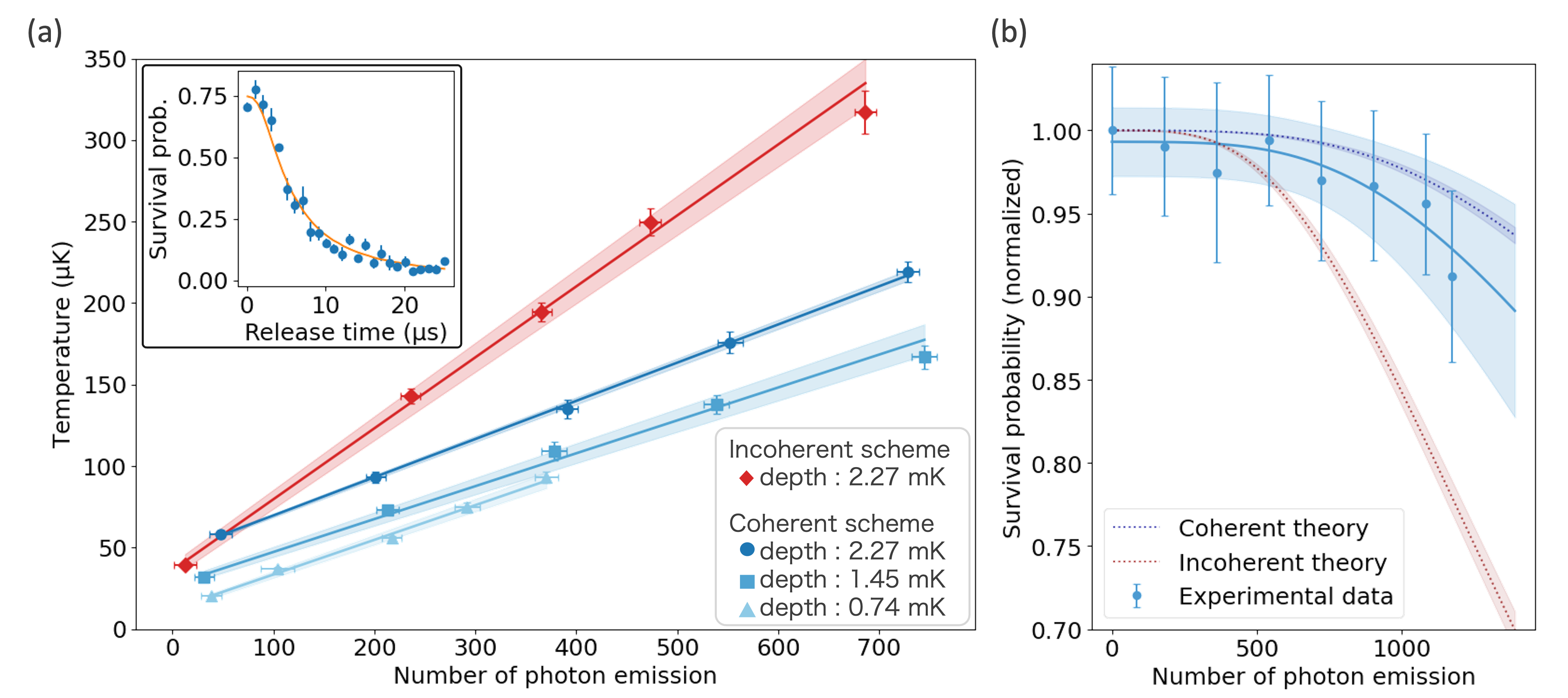}
  \caption{Heating rates during imaging with coherent and incoherent excitation schemes. 
  (a) Evolution of the atomic temperature as a function of the number of pulses, measured via the release-and-recapture method. Number of photon emission is extracted from detected counts by accounting for photon collection efficiency. The data with blue and red circles represent the results in the cases of coherent and incoherent excitation at a trap depth of 2.27 mK, respectively. For the coherent excitation scheme, the experiments are also performed at trap depths of 1.45 mK (blue square) and 0.74 mK (blue triangle) to investigate the effect of DFF. Inset: A representative measurement with the release-and-recapture method, where the survival probability is plotted against the release time to extract the energy distribution. 
  (b) Survival probability normalized to unity in the absence of imaging pulses as a function of emitted photon counts with the coherent excitation scheme at a trap depth of 1.45 mK. The data are represented by blue circles, fitted with Eq.~(\ref{eq:survival_prob}), shown in solid blue line. As a reference, the dashed blue (red) line represents the theoretical heating limits for the coherent (incoherent) schemes.
  In (a) and (b), the error bars indicate the 1-$\sigma$ standard error of the mean.}
  \label{fig:heating_rate}
\end{figure*}

Figure~\ref{fig:heating_rate}(a) shows the results of the temperature measurements with a release-and-recapture method for different total imaging durations, plotted as a function of the number of the spontaneously emitted photons, both for the coherent and incoherent excitation schemes.
The linear dependence of the temperature increase, namely heating, as a function of the photon counts is clearly observed for both methods with different slopes, revealing the different heating rates ($\it{heating~per~photon}$) of atoms for the coherent and incoherent excitation schemes with 234.3(3.4) nK/photon and 435(16) nK/photon, respectively. 
The ratio of the heating rates, 1.9(1), is close to 2, consistent with the expectation.

In order to investigate the possible heating effect known as dipole force fluctuation (DFF) heating~\cite{Dorantes2018}, arising from the non-identical trap depths between the ground and excited states of the probe transition, which is found to be problematic in the case of Rubidium-87 atoms~\cite{Dorantes2018}, we also perform the temperature measurements for the coherent excitation scheme under different trap depths of 1.45 mK and 0.75 mK in addition to the standard condition of 2.27 mK.
The result of the measurements, shown in Fig.~\ref{fig:heating_rate}(a), reveals no net dependence of the heating rate on the trap depth, indicating that the DFF heating is irrelevant in our system. 
To quantitatively explain this observation, we measure the differential light shift of the $^1\text{S}_0 \leftrightarrow{} ^1\text{P}_1$ transition due to the tweezer beam to be 3.3(6) MHz/mK or 16 (3)~\% of the ground state light shift. 
Semiclassical calculation of the DFF effect using this value agrees with the observation (see Appendix~\ref{app:DFF}), confirming its minor contribution.

To get further insights on the heating rates of our coherent excitation scheme, we perform two different temperature diagnostics.
First, we measure the dependence of the survival probability on the number of photon emission at the trap depth of 1.45 mK[Fig.~\ref{fig:heating_rate}(b)] .
In this measurement, to prevent atom loss during the readout, low-intensity, long exposure imaging with simultaneous cooling is performed both before and after the coherent scheme pulses to isolate the loss specifically induced by the coherent scheme imaging process.
By fitting the data with the initial temperature and the photons per pulse treated as fixed parameters, and including a normalization coefficient as a free parameter, we obtain a coherent scheme heating rate of 138(33)~nK/photon based on the survival probability dependence on atomic temperature $T$, given by~\cite{Tuchendler2008}
\begin{equation}\label{eq:survival_prob}
    P_{\text{surv}} = 1 - \left( 1 + \eta + \frac{1}{2}\eta^2 \right) \exp\left( -\eta \right),
\end{equation}
where $\eta = U_0 / T$ and $U_0$ is the trap potential depth. 
In addition, we perform an independent evaluation using the adiabatic trap-depth ramp-down method~\cite{Angonga2022, Alt2003, Tuchendler2008}, yielding a heating rate of 135.2(5.3)~nK/photon. The experimental details of the measurement, as well as the underlying principle, is described in Appendix~\ref{app:heating rate evaluation}.

Since these two methods do not rely on critical parameters specific to the release-and-recapture method—such as effective recapture volume and other potential systematic effects~\cite{ref_coolingRR}—the agreement between the observed heating rates and the theoretical prediction of 115 nK/photon provides independent support for the low heating nature of the coherent excitation scheme. Here, the theoretical prediction of 115 nK/photon is derived as $E_{\text{r}} / 3 k_{\text{B}}$ from the equipartition relation $E_{\text{r}} / 3 k_{\text{B}}$  under the 3D harmonic trap, with only the contribution of the spontaneous emission being considered.

\begin{figure}[t]
  \centering
  \includegraphics[width=0.9\linewidth]{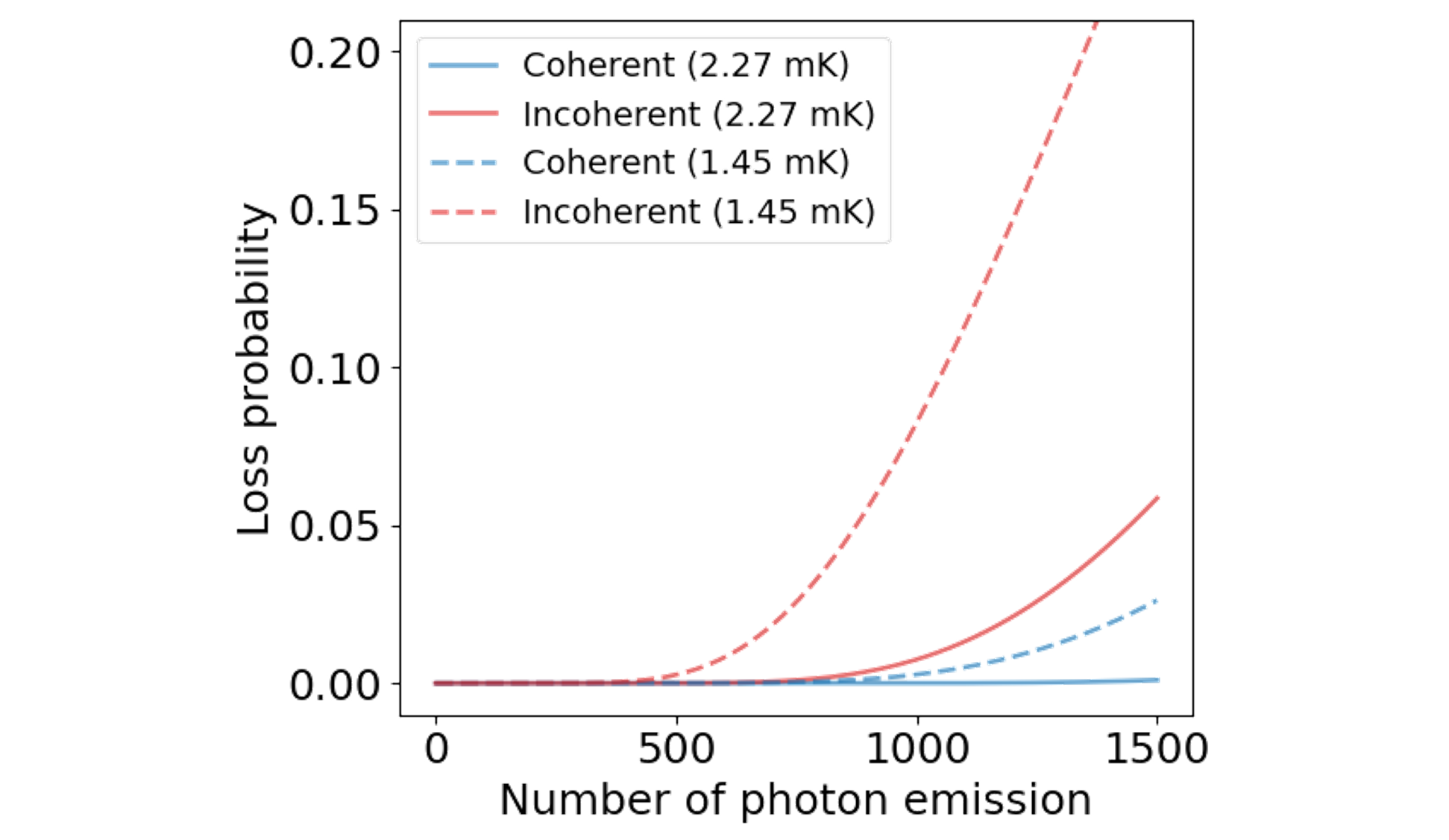}
  \caption{Calculated loss probability as a function of number of photon emission with initial atomic temperature of 30~\textmu K. 
  The blue (red) line represents the result of imaging with coherent (incoherent) excitation scheme.}
  \label{fig:loss_prob}
\end{figure}

Natural consequence of the lower heating rate demonstrated for our imaging scheme is lower loss of atoms in a trap.
Figure~\ref{fig:loss_prob} shows the calculated loss probabilities as a function of number of spontaneous photon emission events during the imaging with coherent and incoherent excitation schemes with initial atomic temperature of 30~\textmu K. The calculation shows the drastic improvement of loss probability for the coherent excitation scheme over the incoherent one.
Even a factor of 2 difference in heating rate results in this large difference in loss probability due to the exponential dependence on the atomic temperature scaled by the trap depth [Eq.~(\ref{eq:survival_prob})].
This advantage in turn relaxes the requirement on the necessary trap depth. It potentially improves the survival probability when limited by off-resonant scattering from the tweezer light~\cite{Falconi2025}. More crucially, the relaxed trap depth leads to a larger scale of atom array sites under a limited power of available trap laser sources.

\section{\label{sec:conclusion}Conclusion}
In conclusion, we have proposed fast, high-fidelity, minimally-destructive imaging scheme for single atoms trapped in an optical tweezer array. 
By applying counter-propagating, alternated $\pi$-pulses on the $^1\text{S}_0 \leftrightarrow{} ^1\text{P}_1$ transition of $^{174}$Yb atoms, we have successfully demonstrated 99.89(5)~\% discrimination fidelity and 98.80(44)~\% survival probability in about 17.6~\textmu s imaging duration, and verified that the heating rate is suppressed by about half compared to the former incoherent excitation scheme.
We expect that improving the light collection efficiency of 3.4~\% by a factor of 3 is realistic by making full use of the high NA 0.6 of our objective.
With this improvement, our method with 300 pulses of less than 6~\textmu s will become the neutral-atom imaging in tweezer arrays as fast as the former state-of-the-arts~\cite{Falconi2025} along with the lower losses, especially useful for quantum-error correction.
The advantage of lower heating rate in our method with coherent excitation could be also applied for imaging of atoms in free space, but with possible reduced spatial resolution due to slightly longer imaging time.

It is straightforward to extend our coherent excitation scheme to other probe transitions of Yb atoms such as $^1\text{S}_0 \leftrightarrow{} ^3\text{P}_1$ and other atomic species.
While high intensity of about $10^4$ mW/cm$^2$ and short pulse of about 1 ns are required in the present case using the $^1\text{P}_1$ state of $^{174}$Yb atoms with a very short lifetime of 5.5 ns, the technical demand for the probe light intensity and pulse width is greatly relaxed for other cases due to longer lifetimes, at the expense of the longer imaging time.
In the case of the $^3\text{P}_1$ state with a lifetime of about 870 ns~\cite{Golub1988} of $^{174}$Yb atoms, we expect similarly high imaging fidelity and survival probability by employing counter-propagating, alternated pulses for sub-millisecond imaging time, provided that the potential of the NA 0.6 optics is sufficiently leveraged.
We envision a dual-isotope-tweezer array platform of Yb atoms would have great benefit of such shorter imaging times and even smaller cross-talks between the isotopes due to the narrow-linewidth and larger isotope shift, over the previous implementation~\cite{Nakamura2024} using the $^1\text{S}_0 \leftrightarrow{} ^1\text{P}_1$ transition for imaging.

The heating is intrinsically linked to the loss associated with imaging, and its reduction offers several key experimental advantages such as the use of shallower traps while maintaining the survival probability, thereby reducing the laser power required per tweezer site, and thus enhancing the scalability of the atom array. 
Conversely, at a fixed trap depth, the loss probability due to heating is expected to decrease exponentially as heating is reduced.
Our method featured with these advantages offers an alternative approach for atom imaging in a variety of quantum science platforms.

\begin{acknowledgments}
We acknowledge Toshihiko Shimasaki for his contridution in earlier stage of this work. We also thank A. Muzi Falconi for insightful discussions. This work was supported by Grants-in-Aid for Scientific Research of JSPS (No. JP22K20356, JP26H00385 and JP26K00639), JST CREST (No. JPMJCR1673 and No. JPMJCR23I3), MEXT Quantum Leap Flagship Program (MEXT Q-LEAP) Grant No. JPMXS0118069021, JST Moonshot R\&D (Grants No. JPMJMS2268, No. JPMJMS2269 and JPMJMS256D), JST ASPIRE (No. JPMJAP24C2), JST PRESTO (No. JPMJPR23F5), the Matsuo Foundation, and JST SPRING (Grant No. JPMJSP2110). R.Y. acknowledges support from the JSPS (KAKENHI Grant No. 25KJ1489).
\end{acknowledgments}


\appendix

\section{Optical setup for short and intense probe beams}
\label{app:optical setp}
\begin{figure*}[t]
  \centering
  \includegraphics[width=0.9\linewidth]{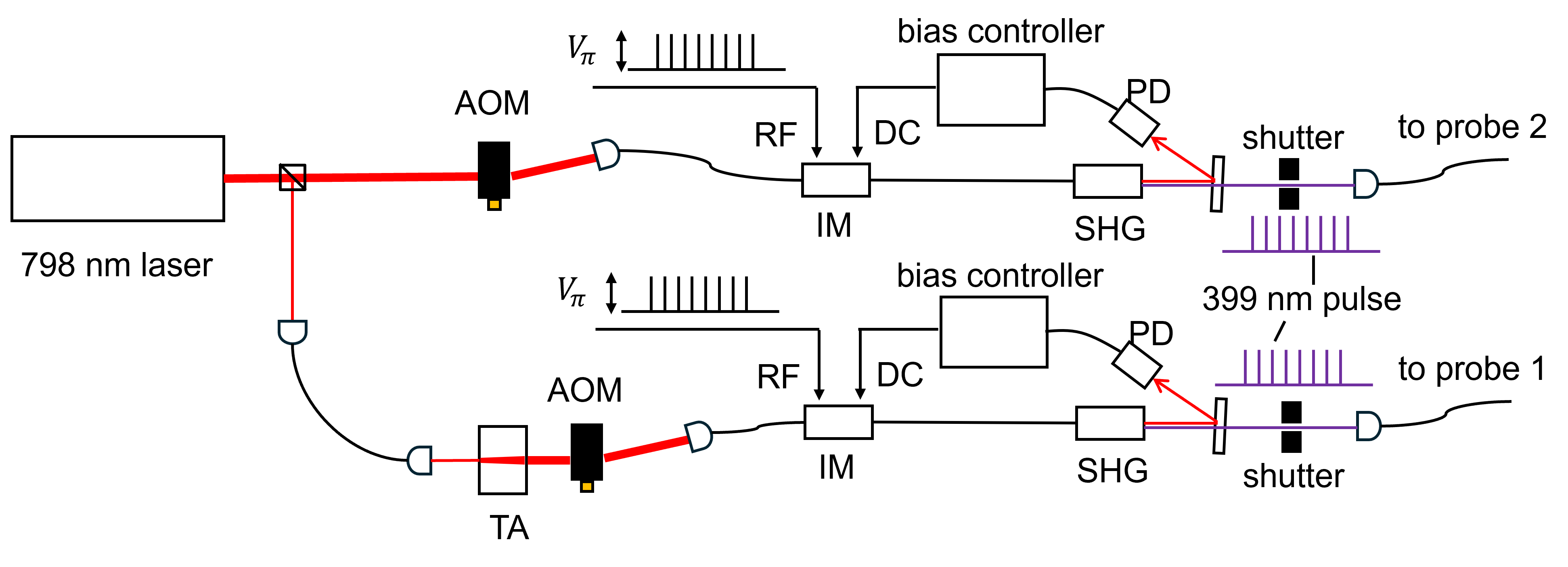}
  \caption{Optical setup for the short and intense probe beams. See text for the basic configuration. The Intensity Modulators (IM) generate the sub-nano second pulses for 798 nm light through the pulse input with the voltage $V_{\pi}$ corresponding to the $\pi$-pulse from the RF ports, whereas the DC ports are used for the bias control to minimize leakage light. The shutters are placed before the input for the optical fibers. AOM, TA, RF, DC, SHG, and PD represents acousto-optic modulators, tapered-amplifiers, radio-frequency, direct current, second-harmonic-generation, and photo-detectors, respectively.}
  \label{fig: opticalsetup}
\end{figure*}
The whole optical setup for generating short and intense probe beams is depicted in Fig.~\ref{fig: opticalsetup}. 
We use a common seed light at 798 nm of 1.7 W from a taper-amplifier (TA) system (TA Pro, Toptica Inc.), the frequency of which is locked to a wavemeter. 
We split the seed 798 nm light into two beams, each of which passes through respective acousto-optic modulator (AOM), EO intensity modulator, and SHG module, with additional TA for one beam. 
Note that we pulse the beams at 798 nm before the EOM to avoid the damage of the EOM and degradation of the SHG module.
At the input of the EO intensity modulator, the 798 nm pulse has a peak power above 500 mW with a low duty cycle.

The output of each SHG module is delivered to the atomic chamber via single-mode polarization maintaining fibers.
We obtain a peak power of up to around 9 mW after the fibers.
Bias voltage on the EO intensity modulator is controlled for a long-term stability.
We use a weak continuous 798 nm light for the bias control. 
The 798 nm light power after the modulator is monitored and 
minimized with a digital servo controller (DSC1, Thorlabs Inc.).
For taking fast single shot imaging in a single experimental run, 
we just turn off the continuous light a few ms before the imaging.
The bias voltage does not change during the interruption of servo control shorter than the servo time constant.
We implement sample and hold for imaging performance evaluation, where two consecutive fast imaging shots are necessary.
The $1/e^2$ waist diameters (horizontal, vertical) of the beams at the atomic position are (140, 134)~\textmu m and (140, 122)~\textmu m, respectively.
The Rabi frequency reaches $2\pi \times 500$ MHz when the beam power at the atom position is 2.5 mW.

We note that there is weak leakage light from the intensity modulator, which can heat atoms.
The extinction ratios of the two intensity modulators are monitored during the experiment and are less than 0.2~\% and around 1~\%, respectively. 
We also observe deterioration of the extinction ratio after irradiating the modulator for a long time. 
While the extinction ratios at 399 nm are higher than these values due to the nonlinear SHG process, we, however, observe heating by the leakage light.
This is reasonable because we use the peak pulse intensity much higher than the saturation intensity.

\section{Evaluation of dipole force fluctuation heating}
\label{app:DFF}
In fluorescence imaging of Rubidium-87, 
it was shown that not only photon recoil but also dipole force fluctuation (DFF) induced heating~\cite{Dorantes2018}. 
Recently, fast imaging of Yb atoms are performed in 532 nm tweezer at a magic-condition~\cite{Falconi2025}. 
Although we use 532 nm tweezer, the polarization condition is not the same as that in~\cite{Falconi2025} and the magic condition is not fulfilled in our experiment.

\begin{figure}[t]
  \centering
  \includegraphics[width=0.9\linewidth]{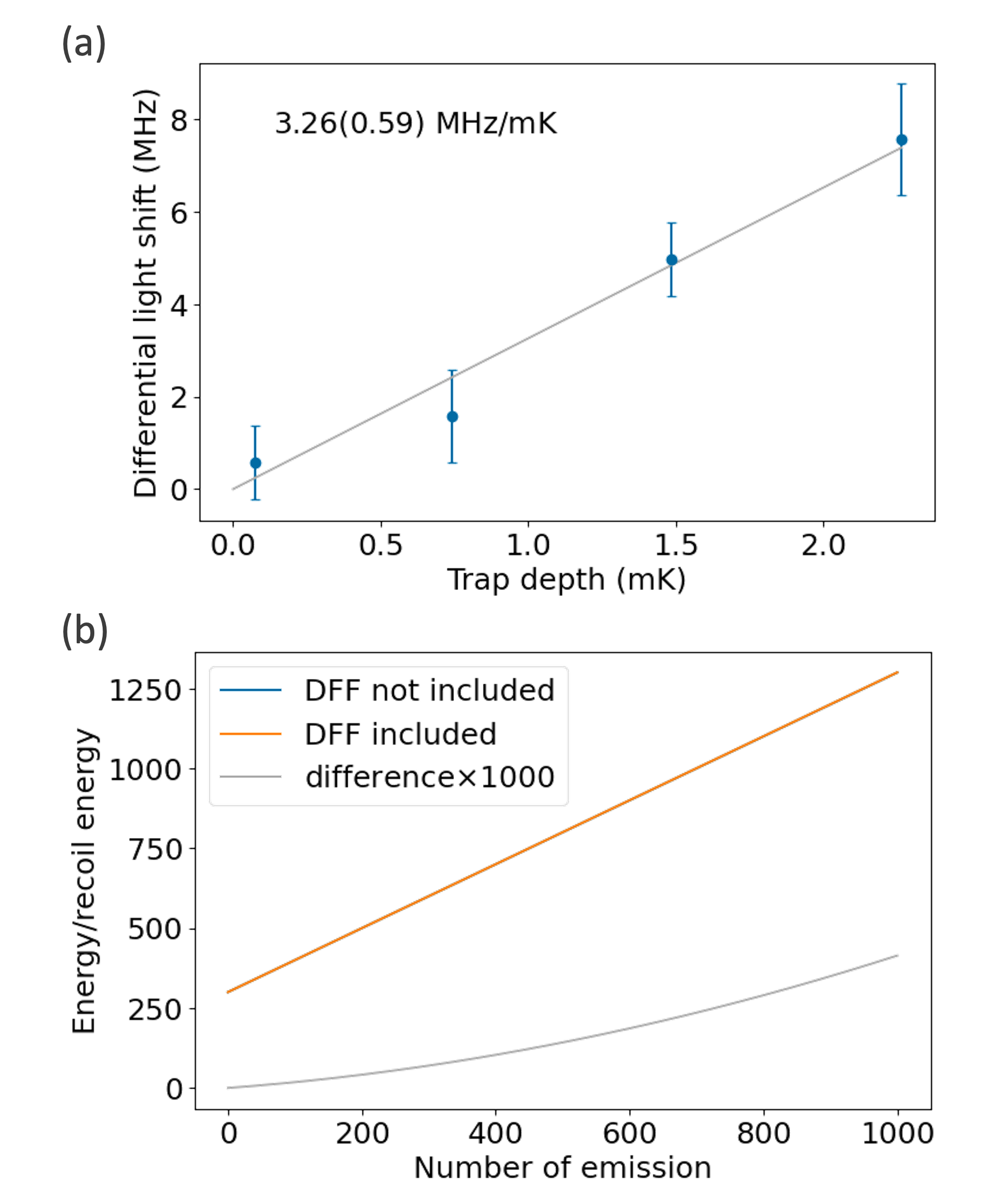}
  \caption{Investigation of the heating due to DFF. 
  (a) Trap depth dependence of the transition frequency for the $^1\text{S}_0 (m_J = 0) \leftrightarrow {}^1\text{P}_1 (m_J = 0)$  line in $^{174}$Yb, measured in a tweezer with a linear polarization angled 45 degrees from the magnetic field. From the linear fit, the differential light shift relative to the trap depth is determined to be approximately 3.3(0.6) MHz/mK. The error bars represent the 1-$\sigma$ standard error of the mean. (b) Calculated heating rate with (orange) and without (blue) DFF. The difference between these two cases, magnified by a factor of 1000 for visual clarity, is shown in gray.}
  \label{fig:DFF}
\end{figure}

Nevertheless, our measurement shows that the heating rate in our experiment is independent of the trap depth, as described in the main text. 
Here we evaluate the DFF effect in our imaging to check the validity of this result.
First, we perform the measurement of the differential light shift of the $^1\text{S}_0 (m_J = 0) \leftrightarrow {}^1\text{P}_1 (m_J = 0)$ transition in a tweezer with a linear polarization angled 45 degrees from the magnetic field, which is shown in Fig.~\ref{fig:DFF}(a), resulting in the determination as 3.3(0.6) MHz/mK.
Next, using this value, we present a classical toy model of our situation. 
We consider a motion in an one-dimentional harmonic trap with the ground and excited state trap frequencies $\omega_g$ and $\omega_e$, respectively. 
Let us consider the energy change during one cycle of excitation at time $t=0$ and de-excitation at time $\Delta t$. 
We denote the initial atom position and velocity as $x_0$ and $v_0$, respectively. 
The atom position and velocity in the excited state at $t$ can be given by
\begin{align}
x(t) &= x_0\cos\left(\omega_e t\right) + \frac{v_0}{\omega_e}\sin\left(\omega_et \right) \notag \\
&= x_0 + v_0t - \frac12 (\omega_e t)^2x_0 + \mathcal{O}(t^3),
\end{align} and
\begin{align}
v(t) &= -\omega_e x_0 \sin\left(\omega_e t\right) + v_0\cos\left(\omega_e t\right)\notag \\
&= v_0 - x_0\omega_e^2 t - \frac12 (\omega_e t)^2v_0 + \mathcal{O}(t^3),
\end{align} respectively. Higher order terms in $t$ are neglected here since $\gamma_1$ is two orders of magnitude larger than the trap frequencies $\omega_{g,e}$.
The total energy after the de-excitation is
\begin{equation}
    E = E_0 + \frac{1}{2}m (\omega_g^2 - \omega_e^2)(v_0^2 -x_0^2 \omega_e^2)\Delta t^2,
\end{equation}
where $E_0 := \frac{1}{2}m\omega_g^2 x_0^2 + \frac{1}{2}mv_0^2$ is the total energy before the excitation.

Assuming the random excitation, the oscillation phase at the time of excitation is also random and we obtain a relation of the energy after $n$-th excitation cycle, $E_n$, as
\begin{equation}\label{eq: energy relation}
    E_{n+1} = \left[1 + \frac{1}{2}(1-\alpha)^2 \left(\frac{\omega_g}{\gamma_1}\right)^2 \right]E_n + E_{\mathrm{r}},
\end{equation}
where $\alpha:=\omega_e^2/\omega_g^2$ and $\Delta t$ is replaced by $1/\gamma_1$. 
The recoil heating is included here.
We note that Eq.~(\ref{eq: energy relation}) with $\alpha=0$ coincides with 
Eq. (5) in~\cite{Dorantes2018}, which was derived under the assumption of a flat excited-state potential.

The solution of Eq.~(\ref{eq: energy relation}) is
\begin{equation} \label{eq: En}
    E_n = E_0 a^n +  E_{\mathrm{}} \frac{a^n-1}{a-1},
\end{equation}
where
\begin{equation}
    a = 1 + \frac{1}{2}(1-\alpha)^2 \left(\frac{\omega_g}{\gamma_1}\right)^2.
\end{equation}

We plot $E_n$ given by Eq.~\ref{eq: En} in Fig.~\ref{fig:DFF} with (orange) and without (blue) DFF, using the experimental parameter $\alpha = 0.84$. 
To illustrate the difference between these two cases, not clear in this plot, we also plot the difference magnified by a factor of 1000, shown in grey. 
These results indicate that DFF heating has a negligible contribution to the total heating in our system.
The small DFF heating is a result of large $\gamma_1$ compared to the trapping frequency $\omega_g$.

\section{Experimental details of incoherent excitation scheme}
\label{app:incoherent}
\begin{figure*}[t]
  \includegraphics[width=0.9\linewidth]{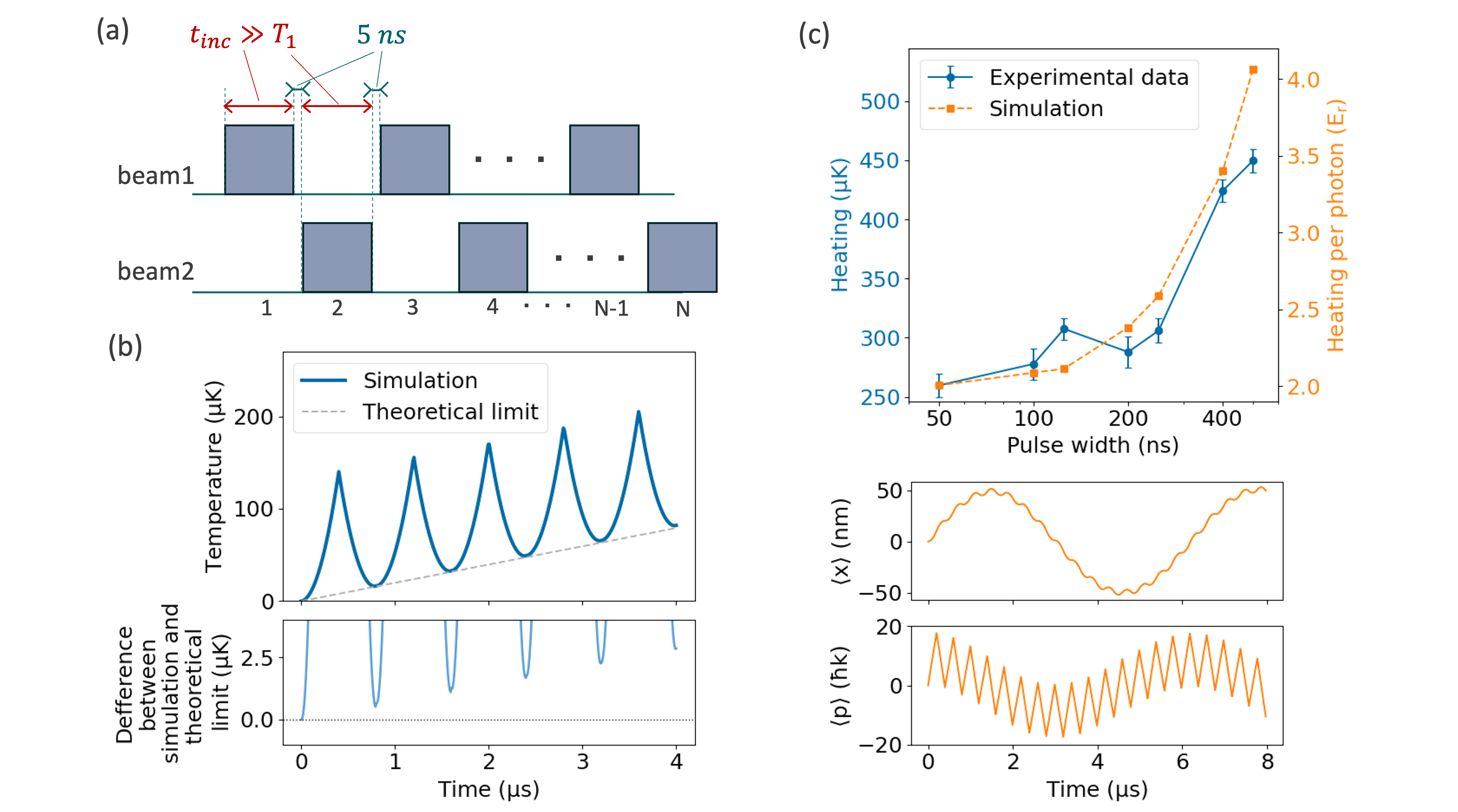}
  \caption{Systematic investigation of the imaging scheme with incoherent excitation. 
  (a) Pulse sequence used for the imaging scheme with incoherent excitation. 
  (b) Numerical simulation of the heating effect including that induced by leakage light (solid lines). The theoretical limit in the absence of leakage light is shown by dashed lines. The lower panel shows the difference between the simulation and the theoretical limit.
  (c) Pulse width dependence of the heating, measured while keeping the total irradiation time constant. The blue circles and the error bars represent the experimental results for the measured heating and the 1-$\sigma$ standard error of fitting, respectively. The orange squares show the results of the semiclassical simulation of heating per photon, accounting for atomic motions within the optical trap. The bottom two panels illustrate the oscillatory trajectories obtained from the simulation.}
  \label{fig:incoherent}
\end{figure*}
The present section details the experimental setup and methodology of the incoherent excitation scheme, as outlined in the main text.
In this work, highly saturated pulses with $I \approx 40~I_{\text{sat}}$ and sufficiently longer width than $T_1$, are alternately applied from opposite directions at a pulse interval of 5 ns (see Fig.~\ref{fig:incoherent}(a)).

To investigate possible heating that arises from intensity imbalances between the two beams, we intentionally introduce an asymmetry by setting the intensity of one beam to $I \approx 60~I_{\text{sat}}$. This measurement is then repeated with the intensity ratio of 2 beams inverted. Both configurations result in no net difference in atomic temperatures from those obtained under the symmetric condition ($I \approx 40~I_{\text{sat}}$), indicating no great role of the intensity asymmetry in heating.

Furthermore, we perform a simulation in order to evaluate the possible effects of leakage light mentioned in the experimental setup section (Appendix~\ref{app:optical setp}).
This simulation employs QuTiP~\cite{qutip2} to simulate the time evolution of momentum (and the resulting atomic temperature) under the following  Hamiltonian~\cite{Su2025}:
\begin{equation}
    H = \frac{\Omega_1}{2} \sum_{p} |p, g\rangle \langle p + 1, e| + \frac{\Omega_2}{2} \sum_{p} |p - 1, e\rangle \langle p, g| + \text{h.c.}
\end{equation}
where $\Omega_{1,2}$ is the Rabi frequency and the index 1 and 2 correspond to the two opposing beams.  
The simulation model employs a sequence of alternated counter-propagating 400 ns pulses. Due to the presence of leakage light, the Rabi frequencies $\Omega_1$ and $\Omega_2$ undergo periodic switching between their peak and leakage-level intensities.
In our simulation, the system acquires momentum along one direction during the initial 400 ns pulse, leading to a local maximum in temperature, as can be seen in Fig.~\ref{fig:incoherent}(b). This momentum is subsequently cancelled by the following 400 ns pulse from the opposite direction, causing the temperature to reach a local minimum. This sequence repeats for a number of pulses, showing a local minimum after an even number of pulses. In the absence of leakage light from the opposing beam, the minimum temperature is solely governed by spontaneous emission and fluctuations in the number of absorbed photons. These processes correspond to the theoretical incoherent scheme heating limit of 19.8~\textmu K/~\textmu s, represented by the dashed line in Fig.~\ref{fig:incoherent}(b).

Next, we incorporate the actual experimental conditions into the simulation. The peak power during the pulse is approximately 40 $I_{\text{sat}}$.
 Regarding the leakage light, fluctuations of up to twofold have been observed over several days. While experimental estimates based on fluorescence intensity suggest a value of 0.04 $I_{\text{sat}}$, we adopt a conservative estimate of 0.4 $I_{\text{sat}}$ for the simulation, the results of which are represented by solid line in Fig.~\ref{fig:incoherent}(b). Notably, even with this tenfold overestimation of the leakage intensity, the additional heating accounts merely for 3.5~\% of the total heating rate.

In addition, with the total imaging duration of 8~\textmu s kept constant, we scan the width of each pulse and observe additional heating for longer pulse widths.
As shown in the upper panel of Fig.~\ref{fig:incoherent}(c), we find that the additional heating remains nearly minimal as long as the pulse width is kept below 100 ns.  
Therefore, unless otherwise specified, the measurements in this work are conducted with the pulse width less than 100 ns.

To investigate the underlying mechanism, we perform semiclassical simulations of an atomic motion in an optical tweezer.
In this model, $^{174}\text{Yb}$ atoms are placed at the center of a 1-D Gaussian potential and subjected to alternated pulses from counter-propagating directions. 
Note that this simulation omits coherent internal or momentum state evolution to isolate the effect of spatial dynamics. 
As previously discussed, a single pulse increases the atomic momentum by $\hbar k N_{\text{scat}}$, leading to a quadratic growth in kinetic energy proportional to $(\hbar k N_{\text{scat}})^2$ (Fig.~\ref{fig:incoherent}(b)), where $ N_{\text{scat}}$ is the number of scattering events during a single pulse. 
In free space, this momentum bias would, on average, be canceled by the subsequent counter-propagating pulse; thus, after an even number of pulses, the only residual heating would stem from spontaneous emission and scattering number fluctuations.
However, our simulations suggest that within a trap, this momentum bias cancellation
is not guaranteed (Fig.~\ref{fig:incoherent}(c)). 
This behavior can be interpreted in the context of the trap oscillation phase. 
At the beginning of the pulse sequence, the atom’s momentum repeatedly shuttles between $0$ and $\hbar k N_{\text{scat}}$, effectively acquiring a ”time-averaged momentum” of
$\hbar k N_{\text{scat}}/2$ that initiates an oscillatory movement. 
Although the total number of pulses is even, if the final pulse is applied at a time when the time-averaged momentum has reversed its original sign, the momentum gained by this pulse will not be cancelled out, but will instead give rise to a momentum bias. Consequently, as the pulse width increases, this momentum bias by the final pulse grows, leading to the drastic increase in kinetic energy.
Using the simulation described above, we plot the simulated heating as a function
of the number of pulses, while keeping the total irradiation time fixed at 8~\textmu s, represented by orange squares plotted with respect to the right axis in (Fig.~\ref{fig:incoherent}(c)).
From the trap frequency in our system, the additional heating due to oscillatory motion
is assumed to be approximately maximized at a total pulse duration of 8.2~\textmu s. 
Thus, our experimental conditions likely represent the regime where this effect is most pronounced.
For visual clarity, the simulation results are scaled to match the experimental data at a pulse width of 50 ns.
The simulation roughly reproduces the scaling behavior of the heating rate observed experimentally, suggesting the possibility that the observed pulse-width dependence is indeed rooted in the atomic oscillation within the trap. 
Furthermore, our simulation predicts that such oscillatory motion can be mitigated by a
sequence that starts and ends with a pulse of half the standard duration, thereby maintaining the time-averaged momentum near zero. 
Such a strategy was employed in~\cite{Falconi2025}, where a 400 ns pulse train was sandwiched by 200 ns pulses. 
In their work, the reported heating rate almost reached the theoretical limit of incoherent scheme.

\section{Heating rate evaluation by adiabatic lowering method}
\label{app:heating rate evaluation}
The adiabatic lowering method involves adiabatically lowering the tweezer potential depth to induce atomic loss, enabling estimation of the energy of atoms that escape the trap~\cite{Angonga2022, Alt2003, Tuchendler2008}. For simplicity, we first consider a one-dimensional model to describe the underlying physics. The action of an atom with energy $E_i$ in a trap of depth $U_i$ is given by $S(E_i, U_i) = \int_{-x_{\mathrm{max}}}^{x_{\mathrm{max}}} \sqrt{2m(E_i - U_ig(x)))} dx$, where $g(x)$ is the normalized potential profile, and $x_{\mathrm{max}}$ is the turning point of the oscillatory motion in the trap. 
An atom escapes the trap when its energy exceeds the potential edge ($E_{\mathrm{esc}} = U_{\mathrm{esc}}$), which defines the mapping between the initial energy $E_i$ and the escape depth $U_{\mathrm{esc}}$ through $S(E_i, U_i) = S(U_{\mathrm{esc}}, U_{\mathrm{esc}})$. By numerically evaluating this relation for the radial direction of our potential trap, the observed survival probability as a function of $U_{\mathrm{esc}}$ is converted into a cumulative energy distribution, $P(E_i) = \int_{0}^{E_i} f(E) dE$. Fitting this distribution with the 3D boltzman distribution $f(E) = \frac{1}{2(k_B T)^3} E^2 e^{-E/k_B T}$
yields the atomic temperature $T$.

In our experiment, the ramp-down, ramp-up, and holding times in the shallow potential are all set to 20 ms. These durations are sufficiently longer than the oscillation period at experimentally relevant trap frequencies of $>$3.7 kHz .
The temperature is evaluated using this method for pulse numbers $N \in \{0, 400, 800, 1200, 1600\}$, followed by linear fitting, yielding a heating rate of coherent scheme 135.2(5.3) nK/photon.

\bibliography{refs}

\begin{thebibliography}{42}%
\makeatletter
\providecommand \@ifxundefined [1]{%
 \@ifx{#1\undefined}
}%
\providecommand \@ifnum [1]{%
 \ifnum #1\expandafter \@firstoftwo
 \else \expandafter \@secondoftwo
 \fi
}%
\providecommand \@ifx [1]{%
 \ifx #1\expandafter \@firstoftwo
 \else \expandafter \@secondoftwo
 \fi
}%
\providecommand \natexlab [1]{#1}%
\providecommand \enquote  [1]{``#1''}%
\providecommand \bibnamefont  [1]{#1}%
\providecommand \bibfnamefont [1]{#1}%
\providecommand \citenamefont [1]{#1}%
\providecommand \href@noop [0]{\@secondoftwo}%
\providecommand \href [0]{\begingroup \@sanitize@url \@href}%
\providecommand \@href[1]{\@@startlink{#1}\@@href}%
\providecommand \@@href[1]{\endgroup#1\@@endlink}%
\providecommand \@sanitize@url [0]{\catcode `\\12\catcode `\$12\catcode `\&12\catcode `\#12\catcode `\^12\catcode `\_12\catcode `\%12\relax}%
\providecommand \@@startlink[1]{}%
\providecommand \@@endlink[0]{}%
\providecommand \url  [0]{\begingroup\@sanitize@url \@url }%
\providecommand \@url [1]{\endgroup\@href {#1}{\urlprefix }}%
\providecommand \urlprefix  [0]{URL }%
\providecommand \Eprint [0]{\href }%
\providecommand \doibase [0]{https://doi.org/}%
\providecommand \selectlanguage [0]{\@gobble}%
\providecommand \bibinfo  [0]{\@secondoftwo}%
\providecommand \bibfield  [0]{\@secondoftwo}%
\providecommand \translation [1]{[#1]}%
\providecommand \BibitemOpen [0]{}%
\providecommand \bibitemStop [0]{}%
\providecommand \bibitemNoStop [0]{.\EOS\space}%
\providecommand \EOS [0]{\spacefactor3000\relax}%
\providecommand \BibitemShut  [1]{\csname bibitem#1\endcsname}%
\let\auto@bib@innerbib\@empty
\bibitem [{\citenamefont {Bakr}\ \emph {et~al.}(2009)\citenamefont {Bakr}, \citenamefont {Gillen}, \citenamefont {Peng}, \citenamefont {F{\"o}lling},\ and\ \citenamefont {Greiner}}]{Bakr2009}%
  \BibitemOpen
  \bibfield  {author} {\bibinfo {author} {\bibfnamefont {W.~S.}\ \bibnamefont {Bakr}}, \bibinfo {author} {\bibfnamefont {J.~I.}\ \bibnamefont {Gillen}}, \bibinfo {author} {\bibfnamefont {A.}~\bibnamefont {Peng}}, \bibinfo {author} {\bibfnamefont {S.}~\bibnamefont {F{\"o}lling}},\ and\ \bibinfo {author} {\bibfnamefont {M.}~\bibnamefont {Greiner}},\ }\bibfield  {title} {\bibinfo {title} {A quantum gas microscope for detecting single atoms in a hubbard-regime optical lattice},\ }\href {https://doi.org/10.1038/nature08482} {\bibfield  {journal} {\bibinfo  {journal} {Nature}\ }\textbf {\bibinfo {volume} {462}},\ \bibinfo {pages} {74} (\bibinfo {year} {2009})}\BibitemShut {NoStop}%
\bibitem [{\citenamefont {Sherson}\ \emph {et~al.}(2010)\citenamefont {Sherson}, \citenamefont {Weitenberg}, \citenamefont {Endres}, \citenamefont {Cheneau}, \citenamefont {Bloch},\ and\ \citenamefont {Kuhr}}]{Sherson2010}%
  \BibitemOpen
  \bibfield  {author} {\bibinfo {author} {\bibfnamefont {J.}~\bibnamefont {Sherson}}, \bibinfo {author} {\bibfnamefont {C.}~\bibnamefont {Weitenberg}}, \bibinfo {author} {\bibfnamefont {M.}~\bibnamefont {Endres}}, \bibinfo {author} {\bibfnamefont {M.}~\bibnamefont {Cheneau}}, \bibinfo {author} {\bibfnamefont {I.}~\bibnamefont {Bloch}},\ and\ \bibinfo {author} {\bibfnamefont {S.}~\bibnamefont {Kuhr}},\ }\bibfield  {title} {\bibinfo {title} {Single-atom-resolved fluorescence imaging of an atomic mott insulator},\ }\href {https://doi.org/10.1038/nature09378} {\bibfield  {journal} {\bibinfo  {journal} {Nature}\ }\textbf {\bibinfo {volume} {467}},\ \bibinfo {pages} {68} (\bibinfo {year} {2010})}\BibitemShut {NoStop}%
\bibitem [{\citenamefont {Gross}\ and\ \citenamefont {Bakr}(2021)}]{Gross2021}%
  \BibitemOpen
  \bibfield  {author} {\bibinfo {author} {\bibfnamefont {C.}~\bibnamefont {Gross}}\ and\ \bibinfo {author} {\bibfnamefont {W.~S.}\ \bibnamefont {Bakr}},\ }\bibfield  {title} {\bibinfo {title} {Quantum gas microscopy for single atom and spin detection},\ }\href {https://doi.org/10.1038/s41567-021-01370-5} {\bibfield  {journal} {\bibinfo  {journal} {Nature Physics}\ }\textbf {\bibinfo {volume} {17}},\ \bibinfo {pages} {1316} (\bibinfo {year} {2021})}\BibitemShut {NoStop}%
\bibitem [{\citenamefont {Kim}\ \emph {et~al.}(2016)\citenamefont {Kim}, \citenamefont {Lee}, \citenamefont {Lee}, \citenamefont {Jo}, \citenamefont {Song},\ and\ \citenamefont {Ahn}}]{Kim2016}%
  \BibitemOpen
  \bibfield  {author} {\bibinfo {author} {\bibfnamefont {H.}~\bibnamefont {Kim}}, \bibinfo {author} {\bibfnamefont {W.}~\bibnamefont {Lee}}, \bibinfo {author} {\bibfnamefont {H.-g.}\ \bibnamefont {Lee}}, \bibinfo {author} {\bibfnamefont {H.}~\bibnamefont {Jo}}, \bibinfo {author} {\bibfnamefont {Y.}~\bibnamefont {Song}},\ and\ \bibinfo {author} {\bibfnamefont {J.}~\bibnamefont {Ahn}},\ }\bibfield  {title} {\bibinfo {title} {In situ single-atom array synthesis using dynamic holographic optical tweezers},\ }\href {https://doi.org/10.1038/ncomms13317} {\bibfield  {journal} {\bibinfo  {journal} {Nature Communications}\ }\textbf {\bibinfo {volume} {7}},\ \bibinfo {pages} {13317} (\bibinfo {year} {2016})}\BibitemShut {NoStop}%
\bibitem [{\citenamefont {Barredo}\ \emph {et~al.}(2016)\citenamefont {Barredo}, \citenamefont {de~L{\'e}s{\'e}leuc}, \citenamefont {Lienhard}, \citenamefont {Lahaye},\ and\ \citenamefont {Browaeys}}]{Barredo2016}%
  \BibitemOpen
  \bibfield  {author} {\bibinfo {author} {\bibfnamefont {D.}~\bibnamefont {Barredo}}, \bibinfo {author} {\bibfnamefont {S.}~\bibnamefont {de~L{\'e}s{\'e}leuc}}, \bibinfo {author} {\bibfnamefont {V.}~\bibnamefont {Lienhard}}, \bibinfo {author} {\bibfnamefont {T.}~\bibnamefont {Lahaye}},\ and\ \bibinfo {author} {\bibfnamefont {A.}~\bibnamefont {Browaeys}},\ }\bibfield  {title} {\bibinfo {title} {An atom-by-atom assembler of defect-free arbitrary two-dimensional atomic arrays},\ }\href {https://doi.org/10.1126/science.aah3778} {\bibfield  {journal} {\bibinfo  {journal} {Science}\ }\textbf {\bibinfo {volume} {354}},\ \bibinfo {pages} {1021} (\bibinfo {year} {2016})}\BibitemShut {NoStop}%
\bibitem [{\citenamefont {Endres}\ \emph {et~al.}(2016)\citenamefont {Endres}, \citenamefont {Bernien}, \citenamefont {Keesling}, \citenamefont {Levine}, \citenamefont {Anschuetz}, \citenamefont {Krajenbrink}, \citenamefont {Senko}, \citenamefont {Vuletic}, \citenamefont {Greiner},\ and\ \citenamefont {Lukin}}]{Endres2016}%
  \BibitemOpen
  \bibfield  {author} {\bibinfo {author} {\bibfnamefont {M.}~\bibnamefont {Endres}}, \bibinfo {author} {\bibfnamefont {H.}~\bibnamefont {Bernien}}, \bibinfo {author} {\bibfnamefont {A.}~\bibnamefont {Keesling}}, \bibinfo {author} {\bibfnamefont {H.}~\bibnamefont {Levine}}, \bibinfo {author} {\bibfnamefont {E.~R.}\ \bibnamefont {Anschuetz}}, \bibinfo {author} {\bibfnamefont {A.}~\bibnamefont {Krajenbrink}}, \bibinfo {author} {\bibfnamefont {C.}~\bibnamefont {Senko}}, \bibinfo {author} {\bibfnamefont {V.}~\bibnamefont {Vuletic}}, \bibinfo {author} {\bibfnamefont {M.}~\bibnamefont {Greiner}},\ and\ \bibinfo {author} {\bibfnamefont {M.~D.}\ \bibnamefont {Lukin}},\ }\bibfield  {title} {\bibinfo {title} {Atom-by-atom assembly of defect-free one-dimensional cold atom arrays},\ }\href {https://doi.org/10.1126/science.aah3752} {\bibfield  {journal} {\bibinfo  {journal} {Science}\ }\textbf {\bibinfo {volume} {354}},\ \bibinfo {pages} {1024} (\bibinfo {year} {2016})}\BibitemShut {NoStop}%
\bibitem [{\citenamefont {Kaufman}\ and\ \citenamefont {Ni}(2021)}]{Kaufman2021}%
  \BibitemOpen
  \bibfield  {author} {\bibinfo {author} {\bibfnamefont {A.~M.}\ \bibnamefont {Kaufman}}\ and\ \bibinfo {author} {\bibfnamefont {K.-K.}\ \bibnamefont {Ni}},\ }\bibfield  {title} {\bibinfo {title} {Quantum science with optical tweezer arrays of ultracold atoms and molecules},\ }\href {https://doi.org/10.1038/s41567-021-01357-2} {\bibfield  {journal} {\bibinfo  {journal} {Nature Physics}\ }\textbf {\bibinfo {volume} {17}},\ \bibinfo {pages} {1324} (\bibinfo {year} {2021})}\BibitemShut {NoStop}%
\bibitem [{\citenamefont {Saffman}\ \emph {et~al.}(2010)\citenamefont {Saffman}, \citenamefont {Walker},\ and\ \citenamefont {M\o{}lmer}}]{Saffman2010}%
  \BibitemOpen
  \bibfield  {author} {\bibinfo {author} {\bibfnamefont {M.}~\bibnamefont {Saffman}}, \bibinfo {author} {\bibfnamefont {T.~G.}\ \bibnamefont {Walker}},\ and\ \bibinfo {author} {\bibfnamefont {K.}~\bibnamefont {M\o{}lmer}},\ }\bibfield  {title} {\bibinfo {title} {Quantum information with rydberg atoms},\ }\href {https://doi.org/10.1103/RevModPhys.82.2313} {\bibfield  {journal} {\bibinfo  {journal} {Rev. Mod. Phys.}\ }\textbf {\bibinfo {volume} {82}},\ \bibinfo {pages} {2313} (\bibinfo {year} {2010})}\BibitemShut {NoStop}%
\bibitem [{\citenamefont {Saffman}(2016)}]{Saffman2016}%
  \BibitemOpen
  \bibfield  {author} {\bibinfo {author} {\bibfnamefont {M.}~\bibnamefont {Saffman}},\ }\bibfield  {title} {\bibinfo {title} {Quantum computing with atomic qubits and rydberg interactions: progress and challenges},\ }\href {https://doi.org/10.1088/0953-4075/49/20/202001} {\bibfield  {journal} {\bibinfo  {journal} {Journal of Physics B: Atomic, Molecular and Optical Physics}\ }\textbf {\bibinfo {volume} {49}},\ \bibinfo {pages} {202001} (\bibinfo {year} {2016})}\BibitemShut {NoStop}%
\bibitem [{\citenamefont {Henriet}\ \emph {et~al.}(2020)\citenamefont {Henriet}, \citenamefont {Beguin}, \citenamefont {Signoles}, \citenamefont {Lahaye}, \citenamefont {Browaeys}, \citenamefont {Reymond},\ and\ \citenamefont {Jurczak}}]{Henriet2020}%
  \BibitemOpen
  \bibfield  {author} {\bibinfo {author} {\bibfnamefont {L.}~\bibnamefont {Henriet}}, \bibinfo {author} {\bibfnamefont {L.}~\bibnamefont {Beguin}}, \bibinfo {author} {\bibfnamefont {A.}~\bibnamefont {Signoles}}, \bibinfo {author} {\bibfnamefont {T.}~\bibnamefont {Lahaye}}, \bibinfo {author} {\bibfnamefont {A.}~\bibnamefont {Browaeys}}, \bibinfo {author} {\bibfnamefont {G.-O.}\ \bibnamefont {Reymond}},\ and\ \bibinfo {author} {\bibfnamefont {C.}~\bibnamefont {Jurczak}},\ }\bibfield  {title} {\bibinfo {title} {Quantum computing with neutral atoms},\ }\href {https://doi.org/10.22331/q-2020-09-21-327} {\bibfield  {journal} {\bibinfo  {journal} {{Quantum}}\ }\textbf {\bibinfo {volume} {4}},\ \bibinfo {pages} {327} (\bibinfo {year} {2020})}\BibitemShut {NoStop}%
\bibitem [{\citenamefont {Wintersperger}\ \emph {et~al.}(2023)\citenamefont {Wintersperger}, \citenamefont {Dommert}, \citenamefont {Ehmer}, \citenamefont {Hoursanov}, \citenamefont {Klepsch}, \citenamefont {Mauerer}, \citenamefont {Reuber}, \citenamefont {Strohm}, \citenamefont {Yin},\ and\ \citenamefont {Luber}}]{Wintersperger2023}%
  \BibitemOpen
  \bibfield  {author} {\bibinfo {author} {\bibfnamefont {K.}~\bibnamefont {Wintersperger}}, \bibinfo {author} {\bibfnamefont {F.}~\bibnamefont {Dommert}}, \bibinfo {author} {\bibfnamefont {T.}~\bibnamefont {Ehmer}}, \bibinfo {author} {\bibfnamefont {A.}~\bibnamefont {Hoursanov}}, \bibinfo {author} {\bibfnamefont {J.}~\bibnamefont {Klepsch}}, \bibinfo {author} {\bibfnamefont {W.}~\bibnamefont {Mauerer}}, \bibinfo {author} {\bibfnamefont {G.}~\bibnamefont {Reuber}}, \bibinfo {author} {\bibfnamefont {T.}~\bibnamefont {Strohm}}, \bibinfo {author} {\bibfnamefont {M.}~\bibnamefont {Yin}},\ and\ \bibinfo {author} {\bibfnamefont {S.}~\bibnamefont {Luber}},\ }\bibfield  {title} {\bibinfo {title} {Neutral atom quantum computing hardware: performance and end-user perspective},\ }\href {https://link.springer.com/article/10.1140/epjqt/s40507-023-00190-1} {\bibfield  {journal} {\bibinfo  {journal} {EPJ Quantum Technol.}\ }\textbf {\bibinfo {volume} {10}} (\bibinfo {year} {2023})}\BibitemShut {NoStop}%
\bibitem [{\citenamefont {Saffman}(2025)}]{saffman2025}%
  \BibitemOpen
  \bibfield  {author} {\bibinfo {author} {\bibfnamefont {M.}~\bibnamefont {Saffman}},\ }\href {https://arxiv.org/abs/2505.11218} {\bibinfo {title} {Quantum computing with atomic qubit arrays: confronting the cost of connectivity}} (\bibinfo {year} {2025}),\ \Eprint {https://arxiv.org/abs/2505.11218} {arXiv:2505.11218 [quant-ph]} \BibitemShut {NoStop}%
\bibitem [{\citenamefont {Madjarov}\ \emph {et~al.}(2019)\citenamefont {Madjarov}, \citenamefont {Cooper}, \citenamefont {Shaw}, \citenamefont {Covey}, \citenamefont {Schkolnik}, \citenamefont {Yoon}, \citenamefont {Williams},\ and\ \citenamefont {Endres}}]{Madjarov2019}%
  \BibitemOpen
  \bibfield  {author} {\bibinfo {author} {\bibfnamefont {I.~S.}\ \bibnamefont {Madjarov}}, \bibinfo {author} {\bibfnamefont {A.}~\bibnamefont {Cooper}}, \bibinfo {author} {\bibfnamefont {A.~L.}\ \bibnamefont {Shaw}}, \bibinfo {author} {\bibfnamefont {J.~P.}\ \bibnamefont {Covey}}, \bibinfo {author} {\bibfnamefont {V.}~\bibnamefont {Schkolnik}}, \bibinfo {author} {\bibfnamefont {T.~H.}\ \bibnamefont {Yoon}}, \bibinfo {author} {\bibfnamefont {J.~R.}\ \bibnamefont {Williams}},\ and\ \bibinfo {author} {\bibfnamefont {M.}~\bibnamefont {Endres}},\ }\bibfield  {title} {\bibinfo {title} {An atomic-array optical clock with single-atom readout},\ }\href {https://doi.org/10.1103/PhysRevX.9.041052} {\bibfield  {journal} {\bibinfo  {journal} {Phys. Rev. X}\ }\textbf {\bibinfo {volume} {9}},\ \bibinfo {pages} {041052} (\bibinfo {year} {2019})}\BibitemShut {NoStop}%
\bibitem [{\citenamefont {Norcia}\ \emph {et~al.}(2019)\citenamefont {Norcia}, \citenamefont {Young}, \citenamefont {Eckner}, \citenamefont {Oelker}, \citenamefont {Ye},\ and\ \citenamefont {Kaufman}}]{Norcia2019}%
  \BibitemOpen
  \bibfield  {author} {\bibinfo {author} {\bibfnamefont {M.~A.}\ \bibnamefont {Norcia}}, \bibinfo {author} {\bibfnamefont {A.~W.}\ \bibnamefont {Young}}, \bibinfo {author} {\bibfnamefont {W.~J.}\ \bibnamefont {Eckner}}, \bibinfo {author} {\bibfnamefont {E.}~\bibnamefont {Oelker}}, \bibinfo {author} {\bibfnamefont {J.}~\bibnamefont {Ye}},\ and\ \bibinfo {author} {\bibfnamefont {A.~M.}\ \bibnamefont {Kaufman}},\ }\bibfield  {title} {\bibinfo {title} {Seconds-scale coherence on an optical clock transition in a tweezer array},\ }\href {https://doi.org/10.1126/science.aay0644} {\bibfield  {journal} {\bibinfo  {journal} {Science}\ }\textbf {\bibinfo {volume} {366}},\ \bibinfo {pages} {93} (\bibinfo {year} {2019})}\BibitemShut {NoStop}%
\bibitem [{\citenamefont {Finkelstein}\ \emph {et~al.}(2024)\citenamefont {Finkelstein}, \citenamefont {Tsai}, \citenamefont {Sun}, \citenamefont {Schine},\ and\ \citenamefont {Endres}}]{Finkelstein2024}%
  \BibitemOpen
  \bibfield  {author} {\bibinfo {author} {\bibfnamefont {R.}~\bibnamefont {Finkelstein}}, \bibinfo {author} {\bibfnamefont {R.~B.~S.}\ \bibnamefont {Tsai}}, \bibinfo {author} {\bibfnamefont {X.}~\bibnamefont {Sun}}, \bibinfo {author} {\bibfnamefont {N.}~\bibnamefont {Schine}},\ and\ \bibinfo {author} {\bibfnamefont {M.}~\bibnamefont {Endres}},\ }\bibfield  {title} {\bibinfo {title} {Universal quantum operations and ancilla-based read-out for tweezer clocks},\ }\href {https://doi.org/10.1038/s41586-024-08005-8} {\bibfield  {journal} {\bibinfo  {journal} {Nature}\ }\textbf {\bibinfo {volume} {634}},\ \bibinfo {pages} {321} (\bibinfo {year} {2024})}\BibitemShut {NoStop}%
\bibitem [{\citenamefont {Kaubruegger}\ and\ \citenamefont {Kaufman}(2025)}]{kaubruegger2025}%
  \BibitemOpen
  \bibfield  {author} {\bibinfo {author} {\bibfnamefont {R.}~\bibnamefont {Kaubruegger}}\ and\ \bibinfo {author} {\bibfnamefont {A.~M.}\ \bibnamefont {Kaufman}},\ }\href {https://arxiv.org/abs/2512.02202} {\bibinfo {title} {Progress in quantum metrology and applications for optical atomic clocks}} (\bibinfo {year} {2025}),\ \Eprint {https://arxiv.org/abs/2512.02202} {arXiv:2512.02202 [quant-ph]} \BibitemShut {NoStop}%
\bibitem [{\citenamefont {Bloch}\ \emph {et~al.}(2012)\citenamefont {Bloch}, \citenamefont {Dalibard},\ and\ \citenamefont {Nascimb^^c3^^a8ne}}]{Bloch2012}%
  \BibitemOpen
  \bibfield  {author} {\bibinfo {author} {\bibfnamefont {I.}~\bibnamefont {Bloch}}, \bibinfo {author} {\bibfnamefont {J.}~\bibnamefont {Dalibard}},\ and\ \bibinfo {author} {\bibfnamefont {S.}~\bibnamefont {Nascimb^^c3^^a8ne}},\ }\bibfield  {title} {\bibinfo {title} {Quantum simulations with ultracold quantum gases},\ }\href {https://doi.org/10.1038/nphys2259} {\bibfield  {journal} {\bibinfo  {journal} {Nature Physics}\ }\textbf {\bibinfo {volume} {8}},\ \bibinfo {pages} {267^^e2^^80^^93276} (\bibinfo {year} {2012})}\BibitemShut {NoStop}%
\bibitem [{\citenamefont {Gross}\ and\ \citenamefont {Bloch}(2017)}]{Gross2017}%
  \BibitemOpen
  \bibfield  {author} {\bibinfo {author} {\bibfnamefont {C.}~\bibnamefont {Gross}}\ and\ \bibinfo {author} {\bibfnamefont {I.}~\bibnamefont {Bloch}},\ }\bibfield  {title} {\bibinfo {title} {Quantum simulations with ultracold atoms in optical lattices},\ }\href {https://doi.org/10.1126/science.aal3837} {\bibfield  {journal} {\bibinfo  {journal} {Science}\ }\textbf {\bibinfo {volume} {357}},\ \bibinfo {pages} {995} (\bibinfo {year} {2017})}\BibitemShut {NoStop}%
\bibitem [{\citenamefont {Sch\"{a}fer}\ \emph {et~al.}(2020)\citenamefont {Sch\"{a}fer}, \citenamefont {Fukuhara}, \citenamefont {Sugawa}, \citenamefont {Takasu},\ and\ \citenamefont {Takahashi}}]{Schfer2020}%
  \BibitemOpen
  \bibfield  {author} {\bibinfo {author} {\bibfnamefont {F.}~\bibnamefont {Sch\"{a}fer}}, \bibinfo {author} {\bibfnamefont {T.}~\bibnamefont {Fukuhara}}, \bibinfo {author} {\bibfnamefont {S.}~\bibnamefont {Sugawa}}, \bibinfo {author} {\bibfnamefont {Y.}~\bibnamefont {Takasu}},\ and\ \bibinfo {author} {\bibfnamefont {Y.}~\bibnamefont {Takahashi}},\ }\bibfield  {title} {\bibinfo {title} {Tools for quantum simulation with ultracold atoms in optical lattices},\ }\href {https://doi.org/10.1038/s42254-020-0195-3} {\bibfield  {journal} {\bibinfo  {journal} {Nature Reviews Physics}\ }\textbf {\bibinfo {volume} {2}},\ \bibinfo {pages} {411^^e2^^80^^93425} (\bibinfo {year} {2020})}\BibitemShut {NoStop}%
\bibitem [{\citenamefont {Browaeys}\ and\ \citenamefont {Lahaye}(2020)}]{Browaeys2020}%
  \BibitemOpen
  \bibfield  {author} {\bibinfo {author} {\bibfnamefont {A.}~\bibnamefont {Browaeys}}\ and\ \bibinfo {author} {\bibfnamefont {T.}~\bibnamefont {Lahaye}},\ }\bibfield  {title} {\bibinfo {title} {Many-body physics with individually controlled rydberg atoms},\ }\href {https://doi.org/10.1038/s42567-019-0733-z} {\bibfield  {journal} {\bibinfo  {journal} {Nature Physics}\ }\textbf {\bibinfo {volume} {16}},\ \bibinfo {pages} {132^^e2^^80^^93142} (\bibinfo {year} {2020})}\BibitemShut {NoStop}%
\bibitem [{\citenamefont {Ott}(2016)}]{Ott2016}%
  \BibitemOpen
  \bibfield  {author} {\bibinfo {author} {\bibfnamefont {H.}~\bibnamefont {Ott}},\ }\bibfield  {title} {\bibinfo {title} {Single atom detection in ultracold quantum gases: a review of current progress},\ }\href {https://doi.org/10.1088/0034-4885/79/5/054401} {\bibfield  {journal} {\bibinfo  {journal} {Reports on Progress in Physics}\ }\textbf {\bibinfo {volume} {79}},\ \bibinfo {pages} {054401} (\bibinfo {year} {2016})}\BibitemShut {NoStop}%
\bibitem [{\citenamefont {Covey}\ \emph {et~al.}(2019)\citenamefont {Covey}, \citenamefont {Madjarov}, \citenamefont {Cooper},\ and\ \citenamefont {Endres}}]{Covey2019}%
  \BibitemOpen
  \bibfield  {author} {\bibinfo {author} {\bibfnamefont {J.~P.}\ \bibnamefont {Covey}}, \bibinfo {author} {\bibfnamefont {I.~S.}\ \bibnamefont {Madjarov}}, \bibinfo {author} {\bibfnamefont {A.}~\bibnamefont {Cooper}},\ and\ \bibinfo {author} {\bibfnamefont {M.}~\bibnamefont {Endres}},\ }\bibfield  {title} {\bibinfo {title} {2000-times repeated imaging of strontium atoms in clock-magic tweezer arrays},\ }\href {https://doi.org/10.1103/PhysRevLett.122.173201} {\bibfield  {journal} {\bibinfo  {journal} {Phys. Rev. Lett.}\ }\textbf {\bibinfo {volume} {122}},\ \bibinfo {pages} {173201} (\bibinfo {year} {2019})}\BibitemShut {NoStop}%
\bibitem [{\citenamefont {Zhou}\ \emph {et~al.}(2025)\citenamefont {Zhou}, \citenamefont {Duckering}, \citenamefont {Zhao}, \citenamefont {Bluvstein}, \citenamefont {Cain}, \citenamefont {Kubica}, \citenamefont {Wang},\ and\ \citenamefont {Lukin}}]{Zhou2025}%
  \BibitemOpen
  \bibfield  {author} {\bibinfo {author} {\bibfnamefont {H.}~\bibnamefont {Zhou}}, \bibinfo {author} {\bibfnamefont {C.}~\bibnamefont {Duckering}}, \bibinfo {author} {\bibfnamefont {C.}~\bibnamefont {Zhao}}, \bibinfo {author} {\bibfnamefont {D.}~\bibnamefont {Bluvstein}}, \bibinfo {author} {\bibfnamefont {M.}~\bibnamefont {Cain}}, \bibinfo {author} {\bibfnamefont {A.}~\bibnamefont {Kubica}}, \bibinfo {author} {\bibfnamefont {S.-T.}\ \bibnamefont {Wang}},\ and\ \bibinfo {author} {\bibfnamefont {M.~D.}\ \bibnamefont {Lukin}},\ }\bibfield  {title} {\bibinfo {title} {Resource analysis of low-overhead transversal architectures for reconfigurable atom arrays},\ }in\ \href {https://doi.org/10.1145/3695053.3731039} {\emph {\bibinfo {booktitle} {Proceedings of the 52nd Annual International Symposium on Computer Architecture}}},\ \bibinfo {series and number} {SIGARCH ’25}\ (\bibinfo  {publisher} {ACM},\ \bibinfo {year} {2025})\ p.\ \bibinfo {pages} {1432^^e2^^80^^931448}\BibitemShut {NoStop}%
\bibitem [{\citenamefont {Sunami}\ \emph {et~al.}(2025)\citenamefont {Sunami}, \citenamefont {Goban},\ and\ \citenamefont {Yamasaki}}]{sunami2025}%
  \BibitemOpen
  \bibfield  {author} {\bibinfo {author} {\bibfnamefont {S.}~\bibnamefont {Sunami}}, \bibinfo {author} {\bibfnamefont {A.}~\bibnamefont {Goban}},\ and\ \bibinfo {author} {\bibfnamefont {H.}~\bibnamefont {Yamasaki}},\ }\href {https://arxiv.org/abs/2506.18979} {\bibinfo {title} {Transversal surface-code game powered by neutral atoms}} (\bibinfo {year} {2025}),\ \Eprint {https://arxiv.org/abs/2506.18979} {arXiv:2506.18979 [quant-ph]} \BibitemShut {NoStop}%
\bibitem [{\citenamefont {Webster}\ \emph {et~al.}(2026)\citenamefont {Webster}, \citenamefont {Berent}, \citenamefont {Chandra}, \citenamefont {Hockings}, \citenamefont {Baspin}, \citenamefont {Thomsen}, \citenamefont {Smith},\ and\ \citenamefont {Cohen}}]{webster2026}%
  \BibitemOpen
  \bibfield  {author} {\bibinfo {author} {\bibfnamefont {P.}~\bibnamefont {Webster}}, \bibinfo {author} {\bibfnamefont {L.}~\bibnamefont {Berent}}, \bibinfo {author} {\bibfnamefont {O.}~\bibnamefont {Chandra}}, \bibinfo {author} {\bibfnamefont {E.~T.}\ \bibnamefont {Hockings}}, \bibinfo {author} {\bibfnamefont {N.}~\bibnamefont {Baspin}}, \bibinfo {author} {\bibfnamefont {F.}~\bibnamefont {Thomsen}}, \bibinfo {author} {\bibfnamefont {S.~C.}\ \bibnamefont {Smith}},\ and\ \bibinfo {author} {\bibfnamefont {L.~Z.}\ \bibnamefont {Cohen}},\ }\href {https://arxiv.org/abs/2602.11457} {\bibinfo {title} {The pinnacle architecture: Reducing the cost of breaking rsa-2048 to 100 000 physical qubits using quantum ldpc codes}} (\bibinfo {year} {2026}),\ \Eprint {https://arxiv.org/abs/2602.11457} {arXiv:2602.11457 [quant-ph]} \BibitemShut {NoStop}%
\bibitem [{\citenamefont {Cain}\ \emph {et~al.}(2026)\citenamefont {Cain}, \citenamefont {Xu}, \citenamefont {King}, \citenamefont {Picard}, \citenamefont {Levine}, \citenamefont {Endres}, \citenamefont {Preskill}, \citenamefont {Huang},\ and\ \citenamefont {Bluvstein}}]{cain2026}%
  \BibitemOpen
  \bibfield  {author} {\bibinfo {author} {\bibfnamefont {M.}~\bibnamefont {Cain}}, \bibinfo {author} {\bibfnamefont {Q.}~\bibnamefont {Xu}}, \bibinfo {author} {\bibfnamefont {R.}~\bibnamefont {King}}, \bibinfo {author} {\bibfnamefont {L.~R.~B.}\ \bibnamefont {Picard}}, \bibinfo {author} {\bibfnamefont {H.}~\bibnamefont {Levine}}, \bibinfo {author} {\bibfnamefont {M.}~\bibnamefont {Endres}}, \bibinfo {author} {\bibfnamefont {J.}~\bibnamefont {Preskill}}, \bibinfo {author} {\bibfnamefont {H.-Y.}\ \bibnamefont {Huang}},\ and\ \bibinfo {author} {\bibfnamefont {D.}~\bibnamefont {Bluvstein}},\ }\href {https://arxiv.org/abs/2603.28627} {\bibinfo {title} {Shor's algorithm is possible with as few as 10,000 reconfigurable atomic qubits}} (\bibinfo {year} {2026}),\ \Eprint {https://arxiv.org/abs/2603.28627} {arXiv:2603.28627 [quant-ph]} \BibitemShut {NoStop}%
\bibitem [{\citenamefont {Xue}\ and\ \citenamefont {Covey}(2026)}]{xue2026}%
  \BibitemOpen
  \bibfield  {author} {\bibinfo {author} {\bibfnamefont {T.}~\bibnamefont {Xue}}\ and\ \bibinfo {author} {\bibfnamefont {J.~P.}\ \bibnamefont {Covey}},\ }\href {https://arxiv.org/abs/2605.03951} {\bibinfo {title} {Factoring $2048$ bit rsa integers with a half-million-qubit modular atomic processor}} (\bibinfo {year} {2026}),\ \Eprint {https://arxiv.org/abs/2605.03951} {arXiv:2605.03951 [quant-ph]} \BibitemShut {NoStop}%
\bibitem [{\citenamefont {Miranda}\ \emph {et~al.}(2015)\citenamefont {Miranda}, \citenamefont {Inoue}, \citenamefont {Okuyama}, \citenamefont {Nakamoto},\ and\ \citenamefont {Kozuma}}]{miranda2015}%
  \BibitemOpen
  \bibfield  {author} {\bibinfo {author} {\bibfnamefont {M.}~\bibnamefont {Miranda}}, \bibinfo {author} {\bibfnamefont {R.}~\bibnamefont {Inoue}}, \bibinfo {author} {\bibfnamefont {Y.}~\bibnamefont {Okuyama}}, \bibinfo {author} {\bibfnamefont {A.}~\bibnamefont {Nakamoto}},\ and\ \bibinfo {author} {\bibfnamefont {M.}~\bibnamefont {Kozuma}},\ }\bibfield  {title} {\bibinfo {title} {Site-resolved imaging of ytterbium atoms in a two-dimensional optical lattice},\ }\href {https://doi.org/10.1103/PhysRevA.91.063414} {\bibfield  {journal} {\bibinfo  {journal} {Phys. Rev. A}\ }\textbf {\bibinfo {volume} {91}},\ \bibinfo {pages} {063414} (\bibinfo {year} {2015})}\BibitemShut {NoStop}%
\bibitem [{\citenamefont {Su}\ \emph {et~al.}(2025)\citenamefont {Su}, \citenamefont {Douglas}, \citenamefont {Szurek}, \citenamefont {H{\'e}bert}, \citenamefont {Krahn}, \citenamefont {Groth}, \citenamefont {Phelps}, \citenamefont {Markovi{\'{c}}},\ and\ \citenamefont {Greiner}}]{Su2025}%
  \BibitemOpen
  \bibfield  {author} {\bibinfo {author} {\bibfnamefont {L.}~\bibnamefont {Su}}, \bibinfo {author} {\bibfnamefont {A.}~\bibnamefont {Douglas}}, \bibinfo {author} {\bibfnamefont {M.}~\bibnamefont {Szurek}}, \bibinfo {author} {\bibfnamefont {A.~H.}\ \bibnamefont {H{\'e}bert}}, \bibinfo {author} {\bibfnamefont {A.}~\bibnamefont {Krahn}}, \bibinfo {author} {\bibfnamefont {R.}~\bibnamefont {Groth}}, \bibinfo {author} {\bibfnamefont {G.~A.}\ \bibnamefont {Phelps}}, \bibinfo {author} {\bibfnamefont {O.}~\bibnamefont {Markovi{\'{c}}}},\ and\ \bibinfo {author} {\bibfnamefont {M.}~\bibnamefont {Greiner}},\ }\bibfield  {title} {\bibinfo {title} {Fast single atom imaging for optical lattice arrays},\ }\href {https://doi.org/10.1038/s41467-025-56305-y} {\bibfield  {journal} {\bibinfo  {journal} {Nature Communications}\ }\textbf {\bibinfo {volume} {16}},\ \bibinfo {pages} {1017} (\bibinfo {year} {2025})}\BibitemShut {NoStop}%
\bibitem [{\citenamefont {Muzi~Falconi}\ \emph {et~al.}(2025)\citenamefont {Muzi~Falconi}, \citenamefont {Panza}, \citenamefont {Sbernardori}, \citenamefont {Forti}, \citenamefont {Klemt}, \citenamefont {Abdel~Karim}, \citenamefont {Marinelli},\ and\ \citenamefont {Scazza}}]{Falconi2025}%
  \BibitemOpen
  \bibfield  {author} {\bibinfo {author} {\bibfnamefont {A.}~\bibnamefont {Muzi~Falconi}}, \bibinfo {author} {\bibfnamefont {R.}~\bibnamefont {Panza}}, \bibinfo {author} {\bibfnamefont {S.}~\bibnamefont {Sbernardori}}, \bibinfo {author} {\bibfnamefont {R.}~\bibnamefont {Forti}}, \bibinfo {author} {\bibfnamefont {R.}~\bibnamefont {Klemt}}, \bibinfo {author} {\bibfnamefont {O.}~\bibnamefont {Abdel~Karim}}, \bibinfo {author} {\bibfnamefont {M.}~\bibnamefont {Marinelli}},\ and\ \bibinfo {author} {\bibfnamefont {F.}~\bibnamefont {Scazza}},\ }\bibfield  {title} {\bibinfo {title} {Microsecond-scale high-survival and number-resolved detection of ytterbium atom arrays},\ }\href {https://doi.org/10.1103/n3bg-7yw7} {\bibfield  {journal} {\bibinfo  {journal} {Phys. Rev. Lett.}\ }\textbf {\bibinfo {volume} {135}},\ \bibinfo {pages} {203402} (\bibinfo {year} {2025})}\BibitemShut {NoStop}%
\bibitem [{\citenamefont {Schaibley}\ \emph {et~al.}(2013)\citenamefont {Schaibley}, \citenamefont {Burgers}, \citenamefont {McCracken}, \citenamefont {Steel}, \citenamefont {Bracker}, \citenamefont {Gammon},\ and\ \citenamefont {Sham}}]{Schaibley2013}%
  \BibitemOpen
  \bibfield  {author} {\bibinfo {author} {\bibfnamefont {J.~R.}\ \bibnamefont {Schaibley}}, \bibinfo {author} {\bibfnamefont {A.~P.}\ \bibnamefont {Burgers}}, \bibinfo {author} {\bibfnamefont {G.~A.}\ \bibnamefont {McCracken}}, \bibinfo {author} {\bibfnamefont {D.~G.}\ \bibnamefont {Steel}}, \bibinfo {author} {\bibfnamefont {A.~S.}\ \bibnamefont {Bracker}}, \bibinfo {author} {\bibfnamefont {D.}~\bibnamefont {Gammon}},\ and\ \bibinfo {author} {\bibfnamefont {L.~J.}\ \bibnamefont {Sham}},\ }\bibfield  {title} {\bibinfo {title} {Direct detection of time-resolved rabi oscillations in a single quantum dot via resonance fluorescence},\ }\href {https://doi.org/10.1103/PhysRevB.87.115311} {\bibfield  {journal} {\bibinfo  {journal} {Phys. Rev. B}\ }\textbf {\bibinfo {volume} {87}},\ \bibinfo {pages} {115311} (\bibinfo {year} {2013})}\BibitemShut {NoStop}%
\bibitem [{\citenamefont {Johansson}\ \emph {et~al.}(2013)\citenamefont {Johansson}, \citenamefont {Nation},\ and\ \citenamefont {Nori}}]{qutip2}%
  \BibitemOpen
  \bibfield  {author} {\bibinfo {author} {\bibfnamefont {J.}~\bibnamefont {Johansson}}, \bibinfo {author} {\bibfnamefont {P.}~\bibnamefont {Nation}},\ and\ \bibinfo {author} {\bibfnamefont {F.}~\bibnamefont {Nori}},\ }\bibfield  {title} {\bibinfo {title} {Qutip 2: A python framework for the dynamics of open quantum systems},\ }\href {https://doi.org/10.1016/j.cpc.2012.11.019} {\bibfield  {journal} {\bibinfo  {journal} {Computer Physics Communications}\ }\textbf {\bibinfo {volume} {184}},\ \bibinfo {pages} {1234^^e2^^80^^931240} (\bibinfo {year} {2013})}\BibitemShut {NoStop}%
\bibitem [{\citenamefont {Norcia}\ \emph {et~al.}(2018)\citenamefont {Norcia}, \citenamefont {Young},\ and\ \citenamefont {Kaufman}}]{Norcia2018}%
  \BibitemOpen
  \bibfield  {author} {\bibinfo {author} {\bibfnamefont {M.~A.}\ \bibnamefont {Norcia}}, \bibinfo {author} {\bibfnamefont {A.~W.}\ \bibnamefont {Young}},\ and\ \bibinfo {author} {\bibfnamefont {A.~M.}\ \bibnamefont {Kaufman}},\ }\bibfield  {title} {\bibinfo {title} {Microscopic control and detection of ultracold strontium in optical-tweezer arrays},\ }\href {https://doi.org/10.1103/PhysRevX.8.041054} {\bibfield  {journal} {\bibinfo  {journal} {Phys. Rev. X}\ }\textbf {\bibinfo {volume} {8}},\ \bibinfo {pages} {041054} (\bibinfo {year} {2018})}\BibitemShut {NoStop}%
\bibitem [{\citenamefont {Holman}\ \emph {et~al.}(2026)\citenamefont {Holman}, \citenamefont {Xu}, \citenamefont {Sun}, \citenamefont {Wu}, \citenamefont {Wang}, \citenamefont {Zhu}, \citenamefont {Seo}, \citenamefont {Yu},\ and\ \citenamefont {Will}}]{Holman2026}%
  \BibitemOpen
  \bibfield  {author} {\bibinfo {author} {\bibfnamefont {A.}~\bibnamefont {Holman}}, \bibinfo {author} {\bibfnamefont {Y.}~\bibnamefont {Xu}}, \bibinfo {author} {\bibfnamefont {X.}~\bibnamefont {Sun}}, \bibinfo {author} {\bibfnamefont {J.}~\bibnamefont {Wu}}, \bibinfo {author} {\bibfnamefont {M.}~\bibnamefont {Wang}}, \bibinfo {author} {\bibfnamefont {Z.}~\bibnamefont {Zhu}}, \bibinfo {author} {\bibfnamefont {B.}~\bibnamefont {Seo}}, \bibinfo {author} {\bibfnamefont {N.}~\bibnamefont {Yu}},\ and\ \bibinfo {author} {\bibfnamefont {S.}~\bibnamefont {Will}},\ }\bibfield  {title} {\bibinfo {title} {Trapping of single atoms in metasurface optical tweezer arrays},\ }\href {https://doi.org/10.1038/s41586-025-09961-5} {\bibfield  {journal} {\bibinfo  {journal} {Nature}\ }\textbf {\bibinfo {volume} {649}},\ \bibinfo {pages} {859^^e2^^80^^93865} (\bibinfo {year} {2026})}\BibitemShut {NoStop}%
\bibitem [{\citenamefont {Bergschneider}\ \emph {et~al.}(2018)\citenamefont {Bergschneider}, \citenamefont {Klinkhamer}, \citenamefont {Becher}, \citenamefont {Klemt}, \citenamefont {Z\"urn}, \citenamefont {Preiss},\ and\ \citenamefont {Jochim}}]{Bergschneider2018}%
  \BibitemOpen
  \bibfield  {author} {\bibinfo {author} {\bibfnamefont {A.}~\bibnamefont {Bergschneider}}, \bibinfo {author} {\bibfnamefont {V.~M.}\ \bibnamefont {Klinkhamer}}, \bibinfo {author} {\bibfnamefont {J.~H.~b.}\ \bibnamefont {Becher}}, \bibinfo {author} {\bibfnamefont {R.}~\bibnamefont {Klemt}}, \bibinfo {author} {\bibfnamefont {G.}~\bibnamefont {Z\"urn}}, \bibinfo {author} {\bibfnamefont {P.~M.}\ \bibnamefont {Preiss}},\ and\ \bibinfo {author} {\bibfnamefont {S.}~\bibnamefont {Jochim}},\ }\bibfield  {title} {\bibinfo {title} {Spin-resolved single-atom imaging of $^{6}\mathrm{Li}$ in free space},\ }\href {https://doi.org/10.1103/PhysRevA.97.063613} {\bibfield  {journal} {\bibinfo  {journal} {Phys. Rev. A}\ }\textbf {\bibinfo {volume} {97}},\ \bibinfo {pages} {063613} (\bibinfo {year} {2018})}\BibitemShut {NoStop}%
\bibitem [{ref()}]{ref_coolingRR}%
  \BibitemOpen
  \href@noop {} {}\bibinfo {note} {We note that the measured temperature exhibits a dependence on the cooling applied after the recapture process; specifically, the inclusion of this post-cooling leads to a lower temperature evaluation. This behavior is likely attributed to the mitigation of excess atom loss during the subsequent imaging phase following the release-and-recapture sequence. It should be emphasized that the post-cooling process does not alter the recapture probability itself. All data presented in this work were acquired with this post-cooling protocol.}\BibitemShut {Stop}%
\bibitem [{\citenamefont {Martinez-Dorantes}\ \emph {et~al.}(2018)\citenamefont {Martinez-Dorantes}, \citenamefont {Alt}, \citenamefont {Gallego}, \citenamefont {Ghosh}, \citenamefont {Ratschbacher},\ and\ \citenamefont {Meschede}}]{Dorantes2018}%
  \BibitemOpen
  \bibfield  {author} {\bibinfo {author} {\bibfnamefont {M.}~\bibnamefont {Martinez-Dorantes}}, \bibinfo {author} {\bibfnamefont {W.}~\bibnamefont {Alt}}, \bibinfo {author} {\bibfnamefont {J.}~\bibnamefont {Gallego}}, \bibinfo {author} {\bibfnamefont {S.}~\bibnamefont {Ghosh}}, \bibinfo {author} {\bibfnamefont {L.}~\bibnamefont {Ratschbacher}},\ and\ \bibinfo {author} {\bibfnamefont {D.}~\bibnamefont {Meschede}},\ }\bibfield  {title} {\bibinfo {title} {State-dependent fluorescence of neutral atoms in optical potentials},\ }\href {https://doi.org/10.1103/PhysRevA.97.023410} {\bibfield  {journal} {\bibinfo  {journal} {Phys. Rev. A}\ }\textbf {\bibinfo {volume} {97}},\ \bibinfo {pages} {023410} (\bibinfo {year} {2018})}\BibitemShut {NoStop}%
\bibitem [{\citenamefont {Tuchendler}\ \emph {et~al.}(2008)\citenamefont {Tuchendler}, \citenamefont {Lance}, \citenamefont {Browaeys}, \citenamefont {Sortais},\ and\ \citenamefont {Grangier}}]{Tuchendler2008}%
  \BibitemOpen
  \bibfield  {author} {\bibinfo {author} {\bibfnamefont {C.}~\bibnamefont {Tuchendler}}, \bibinfo {author} {\bibfnamefont {A.~M.}\ \bibnamefont {Lance}}, \bibinfo {author} {\bibfnamefont {A.}~\bibnamefont {Browaeys}}, \bibinfo {author} {\bibfnamefont {Y.~R.~P.}\ \bibnamefont {Sortais}},\ and\ \bibinfo {author} {\bibfnamefont {P.}~\bibnamefont {Grangier}},\ }\bibfield  {title} {\bibinfo {title} {Energy distribution and cooling of a single atom in an optical tweezer},\ }\href {https://doi.org/10.1103/PhysRevA.78.033425} {\bibfield  {journal} {\bibinfo  {journal} {Phys. Rev. A}\ }\textbf {\bibinfo {volume} {78}},\ \bibinfo {pages} {033425} (\bibinfo {year} {2008})}\BibitemShut {NoStop}%
\bibitem [{\citenamefont {Ang'ong'a}\ \emph {et~al.}(2022)\citenamefont {Ang'ong'a}, \citenamefont {Huang}, \citenamefont {Covey},\ and\ \citenamefont {Gadway}}]{Angonga2022}%
  \BibitemOpen
  \bibfield  {author} {\bibinfo {author} {\bibfnamefont {J.}~\bibnamefont {Ang'ong'a}}, \bibinfo {author} {\bibfnamefont {C.}~\bibnamefont {Huang}}, \bibinfo {author} {\bibfnamefont {J.~P.}\ \bibnamefont {Covey}},\ and\ \bibinfo {author} {\bibfnamefont {B.}~\bibnamefont {Gadway}},\ }\bibfield  {title} {\bibinfo {title} {Gray molasses cooling of {$^{39}$K} atoms in optical tweezers},\ }\href {https://doi.org/10.1103/PhysRevResearch.4.013240} {\bibfield  {journal} {\bibinfo  {journal} {Phys. Rev. Research}\ }\textbf {\bibinfo {volume} {4}},\ \bibinfo {pages} {013240} (\bibinfo {year} {2022})}\BibitemShut {NoStop}%
\bibitem [{\citenamefont {Alt}\ \emph {et~al.}(2003)\citenamefont {Alt}, \citenamefont {Schrader}, \citenamefont {Kuhr}, \citenamefont {M\"uller}, \citenamefont {Gomer},\ and\ \citenamefont {Meschede}}]{Alt2003}%
  \BibitemOpen
  \bibfield  {author} {\bibinfo {author} {\bibfnamefont {W.}~\bibnamefont {Alt}}, \bibinfo {author} {\bibfnamefont {D.}~\bibnamefont {Schrader}}, \bibinfo {author} {\bibfnamefont {S.}~\bibnamefont {Kuhr}}, \bibinfo {author} {\bibfnamefont {M.}~\bibnamefont {M\"uller}}, \bibinfo {author} {\bibfnamefont {V.}~\bibnamefont {Gomer}},\ and\ \bibinfo {author} {\bibfnamefont {D.}~\bibnamefont {Meschede}},\ }\bibfield  {title} {\bibinfo {title} {Single atoms in a standing-wave dipole trap},\ }\href {https://doi.org/10.1103/PhysRevA.67.033403} {\bibfield  {journal} {\bibinfo  {journal} {Phys. Rev. A}\ }\textbf {\bibinfo {volume} {67}},\ \bibinfo {pages} {033403} (\bibinfo {year} {2003})}\BibitemShut {NoStop}%
\bibitem [{\citenamefont {Golub}\ \emph {et~al.}(1988)\citenamefont {Golub}, \citenamefont {Bai},\ and\ \citenamefont {Mossberg}}]{Golub1988}%
  \BibitemOpen
  \bibfield  {author} {\bibinfo {author} {\bibfnamefont {J.~E.}\ \bibnamefont {Golub}}, \bibinfo {author} {\bibfnamefont {Y.~S.}\ \bibnamefont {Bai}},\ and\ \bibinfo {author} {\bibfnamefont {T.~W.}\ \bibnamefont {Mossberg}},\ }\bibfield  {title} {\bibinfo {title} {Radiative and dynamical properties of homogeneously prepared atomic samples},\ }\href {https://doi.org/10.1103/PhysRevA.37.119} {\bibfield  {journal} {\bibinfo  {journal} {Phys. Rev. A}\ }\textbf {\bibinfo {volume} {37}},\ \bibinfo {pages} {119} (\bibinfo {year} {1988})}\BibitemShut {NoStop}%
\bibitem [{\citenamefont {Nakamura}\ \emph {et~al.}(2024)\citenamefont {Nakamura}, \citenamefont {Kusano}, \citenamefont {Yokoyama}, \citenamefont {Saito}, \citenamefont {Higashi}, \citenamefont {Ozawa}, \citenamefont {Takano}, \citenamefont {Takasu},\ and\ \citenamefont {Takahashi}}]{Nakamura2024}%
  \BibitemOpen
  \bibfield  {author} {\bibinfo {author} {\bibfnamefont {Y.}~\bibnamefont {Nakamura}}, \bibinfo {author} {\bibfnamefont {T.}~\bibnamefont {Kusano}}, \bibinfo {author} {\bibfnamefont {R.}~\bibnamefont {Yokoyama}}, \bibinfo {author} {\bibfnamefont {K.}~\bibnamefont {Saito}}, \bibinfo {author} {\bibfnamefont {K.}~\bibnamefont {Higashi}}, \bibinfo {author} {\bibfnamefont {N.}~\bibnamefont {Ozawa}}, \bibinfo {author} {\bibfnamefont {T.}~\bibnamefont {Takano}}, \bibinfo {author} {\bibfnamefont {Y.}~\bibnamefont {Takasu}},\ and\ \bibinfo {author} {\bibfnamefont {Y.}~\bibnamefont {Takahashi}},\ }\bibfield  {title} {\bibinfo {title} {Hybrid atom tweezer array of nuclear spin and optical clock qubits},\ }\href {https://doi.org/10.1103/PhysRevX.14.041062} {\bibfield  {journal} {\bibinfo  {journal} {Phys. Rev. X}\ }\textbf {\bibinfo {volume} {14}},\ \bibinfo {pages} {041062} (\bibinfo {year} {2024})}\BibitemShut {NoStop}%
\end{thebibliography}%

\end{document}